\documentclass[longauth]{aa}

\usepackage{graphicx}
\usepackage{txfonts}

\usepackage{subcaption}
\usepackage{gensymb}
\usepackage{arydshln}
\usepackage{tablefootnote}

\usepackage[colorlinks=true,citecolor=blue]{hyperref}

\usepackage{xcolor}
\usepackage{soul}

\usepackage{import}

\begin{document}

    \title{Precise characterisation of HD 15337 with CHEOPS: a laboratory for planet formation and evolution
        \thanks{This article uses data from CHEOPS programme CH\_PR100031.}\fnmsep
        \thanks{Based on observations made with ESO-3.6 m telescope at the La Silla Observatory under programme ID 1102.C-0923.}}    
    
        \author{
N. M. Rosário\inst{\ref{inst:1},\ref{inst:2}} $^{\href{https://orcid.org/0000-0002-8588-6730}{\includegraphics[scale=0.5]{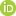}}}$\and 
O. D. S. Demangeon\inst{\ref{inst:1},\ref{inst:2}} $^{\href{https://orcid.org/0000-0001-7918-0355}{\includegraphics[scale=0.5]{figures/orcid.jpg}}}$\and 
S. C. C. Barros\inst{\ref{inst:1},\ref{inst:2}} $^{\href{https://orcid.org/0000-0003-2434-3625}{\includegraphics[scale=0.5]{figures/orcid.jpg}}}$\and 
D. Gandolfi\inst{\ref{inst:3}} $^{\href{https://orcid.org/0000-0001-8627-9628}{\includegraphics[scale=0.5]{figures/orcid.jpg}}}$\and 
J. A. Egger\inst{\ref{inst:4}} $^{\href{https://orcid.org/0000-0003-1628-4231}{\includegraphics[scale=0.5]{figures/orcid.jpg}}}$\and 
L. M. Serrano\inst{\ref{inst:3}} $^{\href{https://orcid.org/0000-0001-9211-3691}{\includegraphics[scale=0.5]{figures/orcid.jpg}}}$\and 
H. P. Osborn\inst{\ref{inst:5},\ref{inst:6}} $^{\href{https://orcid.org/0000-0002-4047-4724}{\includegraphics[scale=0.5]{figures/orcid.jpg}}}$\and 
M. Beck\inst{\ref{inst:7}} $^{\href{https://orcid.org/0000-0003-3926-0275}{\includegraphics[scale=0.5]{figures/orcid.jpg}}}$\and 
W. Benz\inst{\ref{inst:4},\ref{inst:5}} $^{\href{https://orcid.org/0000-0001-7896-6479}{\includegraphics[scale=0.5]{figures/orcid.jpg}}}$\and 
H.-G. Florén\inst{\ref{inst:8}}\and 
P. Guterman\inst{\ref{inst:9},\ref{inst:10}}\and 
T. G. Wilson\inst{\ref{inst:11}} $^{\href{https://orcid.org/0000-0001-8749-1962}{\includegraphics[scale=0.5]{figures/orcid.jpg}}}$\and 
Y. Alibert\inst{\ref{inst:5},\ref{inst:4}} $^{\href{https://orcid.org/0000-0002-4644-8818}{\includegraphics[scale=0.5]{figures/orcid.jpg}}}$\and 
L. Fossati\inst{\ref{inst:12}} $^{\href{https://orcid.org/0000-0003-4426-9530}{\includegraphics[scale=0.5]{figures/orcid.jpg}}}$\and 
M. J. Hooton\inst{\ref{inst:13}} $^{\href{https://orcid.org/0000-0003-0030-332X}{\includegraphics[scale=0.5]{figures/orcid.jpg}}}$\and 
L. Delrez\inst{\ref{inst:14},\ref{inst:15}} $^{\href{https://orcid.org/0000-0001-6108-4808}{\includegraphics[scale=0.5]{figures/orcid.jpg}}}$\and 
N. C. Santos\inst{\ref{inst:1},\ref{inst:2}} $^{\href{https://orcid.org/0000-0003-4422-2919}{\includegraphics[scale=0.5]{figures/orcid.jpg}}}$\and 
S. G. Sousa\inst{\ref{inst:1}} $^{\href{https://orcid.org/0000-0001-9047-2965}{\includegraphics[scale=0.5]{figures/orcid.jpg}}}$\and 
A. Bonfanti\inst{\ref{inst:12}} $^{\href{https://orcid.org/0000-0002-1916-5935}{\includegraphics[scale=0.5]{figures/orcid.jpg}}}$\and 
S. Salmon\inst{\ref{inst:7}} $^{\href{https://orcid.org/0000-0002-1714-3513}{\includegraphics[scale=0.5]{figures/orcid.jpg}}}$\and 
V. Adibekyan\inst{\ref{inst:1}} $^{\href{https://orcid.org/0000-0002-0601-6199}{\includegraphics[scale=0.5]{figures/orcid.jpg}}}$\and
A. Nigioni\inst{\ref{inst:24}} $^{\href{https://orcid.org/0009-0004-5882-6574}{\includegraphics[scale=0.5]{figures/orcid.jpg}}}$\and
J. Venturini\inst{\ref{inst:24}} $^{\href{https://orcid.org/0000-0001-9527-2903}{\includegraphics[scale=0.5]{figures/orcid.jpg}}}$\and
R. Alonso\inst{\ref{inst:16},\ref{inst:17}} $^{\href{https://orcid.org/0000-0001-8462-8126}{\includegraphics[scale=0.5]{figures/orcid.jpg}}}$\and 
G. Anglada\inst{\ref{inst:18},\ref{inst:19}} $^{\href{https://orcid.org/0000-0002-3645-5977}{\includegraphics[scale=0.5]{figures/orcid.jpg}}}$\and 
J. Asquier\inst{\ref{inst:20}}\and 
T. Bárczy\inst{\ref{inst:21}} $^{\href{https://orcid.org/0000-0002-7822-4413}{\includegraphics[scale=0.5]{figures/orcid.jpg}}}$\and 
D. Barrado Navascues\inst{\ref{inst:22}} $^{\href{https://orcid.org/0000-0002-5971-9242}{\includegraphics[scale=0.5]{figures/orcid.jpg}}}$\and 
O. Barragán\inst{\ref{inst:23}}\and 
W. Baumjohann\inst{\ref{inst:12}} $^{\href{https://orcid.org/0000-0001-6271-0110}{\includegraphics[scale=0.5]{figures/orcid.jpg}}}$\and 
T. Beck\inst{\ref{inst:4}}\and 
N. Billot\inst{\ref{inst:24}} $^{\href{https://orcid.org/0000-0003-3429-3836}{\includegraphics[scale=0.5]{figures/orcid.jpg}}}$\and 
F. Biondi\inst{\ref{inst:25}, \ref{inst:26}} $^{\href{https://orcid.org/0000-0002-1337-3653}{\includegraphics[scale=0.5]{figures/orcid.jpg}}}$\and 
X. Bonfils\inst{\ref{inst:27}} $^{\href{https://orcid.org/0000-0001-9003-8894}{\includegraphics[scale=0.5]{figures/orcid.jpg}}}$\and 
L. Borsato\inst{\ref{inst:26}} $^{\href{https://orcid.org/0000-0003-0066-9268}{\includegraphics[scale=0.5]{figures/orcid.jpg}}}$\and 
A. Brandeker\inst{\ref{inst:8}} $^{\href{https://orcid.org/0000-0002-7201-7536}{\includegraphics[scale=0.5]{figures/orcid.jpg}}}$\and 
C. Broeg\inst{\ref{inst:4},\ref{inst:5}} $^{\href{https://orcid.org/0000-0001-5132-2614}{\includegraphics[scale=0.5]{figures/orcid.jpg}}}$\and 
V. Cessa\inst{\ref{inst:4}}\and 
S. Charnoz\inst{\ref{inst:28}} $^{\href{https://orcid.org/0000-0002-7442-491X}{\includegraphics[scale=0.5]{figures/orcid.jpg}}}$\and 
A. Collier Cameron\inst{\ref{inst:11}} $^{\href{https://orcid.org/0000-0002-8863-7828}{\includegraphics[scale=0.5]{figures/orcid.jpg}}}$\and 
Sz. Csizmadia\inst{\ref{inst:29}} $^{\href{https://orcid.org/0000-0001-6803-9698}{\includegraphics[scale=0.5]{figures/orcid.jpg}}}$\and 
P. E. Cubillos\inst{\ref{inst:30},\ref{inst:12}}\and 
M. B. Davies\inst{\ref{inst:31}} $^{\href{https://orcid.org/0000-0001-6080-1190}{\includegraphics[scale=0.5]{figures/orcid.jpg}}}$\and 
M. Deleuil\inst{\ref{inst:9}} $^{\href{https://orcid.org/0000-0001-6036-0225}{\includegraphics[scale=0.5]{figures/orcid.jpg}}}$\and 
A. Deline\inst{\ref{inst:24}}\and 
B.-O. Demory\inst{\ref{inst:5},\ref{inst:4}} $^{\href{https://orcid.org/0000-0002-9355-5165}{\includegraphics[scale=0.5]{figures/orcid.jpg}}}$\and 
D. Ehrenreich\inst{\ref{inst:24},\ref{inst:33}} $^{\href{https://orcid.org/0000-0001-9704-5405}{\includegraphics[scale=0.5]{figures/orcid.jpg}}}$\and 
A. Erikson\inst{\ref{inst:29}}\and 
M. Esposito\inst{\ref{inst:34}}\and 
A. Fortier\inst{\ref{inst:4},\ref{inst:5}} $^{\href{https://orcid.org/0000-0001-8450-3374}{\includegraphics[scale=0.5]{figures/orcid.jpg}}}$\and 
M. Fridlund\inst{\ref{inst:35},\ref{inst:36}} $^{\href{https://orcid.org/0000-0002-0855-8426}{\includegraphics[scale=0.5]{figures/orcid.jpg}}}$\and 
M. Gillon\inst{\ref{inst:14}} $^{\href{https://orcid.org/0000-0003-1462-7739}{\includegraphics[scale=0.5]{figures/orcid.jpg}}}$\and 
M. Güdel\inst{\ref{inst:37}}\and 
M. N. Günther\inst{\ref{inst:20}} $^{\href{https://orcid.org/0000-0002-3164-9086}{\includegraphics[scale=0.5]{figures/orcid.jpg}}}$\and  
Ch. Helling\inst{\ref{inst:12}}\and 
S. Hoyer\inst{\ref{inst:9}} $^{\href{https://orcid.org/0000-0003-3477-2466}{\includegraphics[scale=0.5]{figures/orcid.jpg}}}$\and 
K. G. Isaak\inst{\ref{inst:20}} $^{\href{https://orcid.org/0000-0001-8585-1717}{\includegraphics[scale=0.5]{figures/orcid.jpg}}}$\and 
L. L. Kiss\inst{\ref{inst:38},\ref{inst:39}}\and 
K. W. F. Lam\inst{\ref{inst:29}} $^{\href{https://orcid.org/0000-0002-9910-6088}{\includegraphics[scale=0.5]{figures/orcid.jpg}}}$\and 
J. Laskar\inst{\ref{inst:40}} $^{\href{https://orcid.org/0000-0003-2634-789X}{\includegraphics[scale=0.5]{figures/orcid.jpg}}}$\and 
A. Lecavelier des Etangs\inst{\ref{inst:41}} $^{\href{https://orcid.org/0000-0002-5637-5253}{\includegraphics[scale=0.5]{figures/orcid.jpg}}}$\and 
M. Lendl\inst{\ref{inst:24}} $^{\href{https://orcid.org/0000-0001-9699-1459}{\includegraphics[scale=0.5]{figures/orcid.jpg}}}$\and 
A. Luntzer\inst{\ref{inst:36}}\and 
D. Magrin\inst{\ref{inst:26}} $^{\href{https://orcid.org/0000-0003-0312-313X}{\includegraphics[scale=0.5]{figures/orcid.jpg}}}$\and 
P. F. L. Maxted\inst{\ref{inst:42}} $^{\href{https://orcid.org/0000-0003-3794-1317}{\includegraphics[scale=0.5]{figures/orcid.jpg}}}$\and 
C. Mordasini\inst{\ref{inst:4},\ref{inst:5}}\and 
V. Nascimbeni\inst{\ref{inst:26}} $^{\href{https://orcid.org/0000-0001-9770-1214}{\includegraphics[scale=0.5]{figures/orcid.jpg}}}$\and 
G. Olofsson\inst{\ref{inst:8}} $^{\href{https://orcid.org/0000-0003-3747-7120}{\includegraphics[scale=0.5]{figures/orcid.jpg}}}$\and 
H. L. M. Osborne\inst{\ref{eso}, \ref{inst:43}}\and 
R. Ottensamer\inst{\ref{inst:36}}\and 
I. Pagano\inst{\ref{inst:44}} $^{\href{https://orcid.org/0000-0001-9573-4928}{\includegraphics[scale=0.5]{figures/orcid.jpg}}}$\and 
E. Pallé\inst{\ref{inst:16},\ref{inst:17}} $^{\href{https://orcid.org/0000-0003-0987-1593}{\includegraphics[scale=0.5]{figures/orcid.jpg}}}$\and 
G. Peter\inst{\ref{inst:45}} $^{\href{https://orcid.org/0000-0001-6101-2513}{\includegraphics[scale=0.5]{figures/orcid.jpg}}}$\and 
G. Piotto\inst{\ref{inst:26},\ref{inst:46}} $^{\href{https://orcid.org/0000-0002-9937-6387}{\includegraphics[scale=0.5]{figures/orcid.jpg}}}$\and 
D. Pollacco\inst{\ref{inst:47}}\and 
D. Queloz\inst{\ref{inst:48},\ref{inst:13}} $^{\href{https://orcid.org/0000-0002-3012-0316}{\includegraphics[scale=0.5]{figures/orcid.jpg}}}$\and 
R. Ragazzoni\inst{\ref{inst:26},\ref{inst:46}} $^{\href{https://orcid.org/0000-0002-7697-5555}{\includegraphics[scale=0.5]{figures/orcid.jpg}}}$\and 
N. Rando\inst{\ref{inst:20}}\and 
H. Rauer\inst{\ref{inst:29},\ref{inst:49},\ref{inst:50}} $^{\href{https://orcid.org/0000-0002-6510-1828}{\includegraphics[scale=0.5]{figures/orcid.jpg}}}$\and 
I. Ribas\inst{\ref{inst:18},\ref{inst:19}} $^{\href{https://orcid.org/0000-0002-6689-0312}{\includegraphics[scale=0.5]{figures/orcid.jpg}}}$\and 
G. Scandariato\inst{\ref{inst:44}} $^{\href{https://orcid.org/0000-0003-2029-0626}{\includegraphics[scale=0.5]{figures/orcid.jpg}}}$\and 
D. Ségransan\inst{\ref{inst:31}} $^{\href{https://orcid.org/0000-0003-2355-8034}{\includegraphics[scale=0.5]{figures/orcid.jpg}}}$\and 
A. E. Simon\inst{\ref{inst:4}} $^{\href{https://orcid.org/0000-0001-9773-2600}{\includegraphics[scale=0.5]{figures/orcid.jpg}}}$\and 
A. M. S. Smith\inst{\ref{inst:29}} $^{\href{https://orcid.org/0000-0002-2386-4341}{\includegraphics[scale=0.5]{figures/orcid.jpg}}}$\and 
M. Stalport\inst{\ref{inst:51}}\and 
Gy. M. Szabó\inst{\ref{inst:52},\ref{inst:53}}\and 
N. Thomas\inst{\ref{inst:4}}\and 
S. Udry\inst{\ref{inst:24}} $^{\href{https://orcid.org/0000-0001-7576-6236}{\includegraphics[scale=0.5]{figures/orcid.jpg}}}$\and 
V. Van Eylen\inst{\ref{inst:43}}\and 
V. Van Grootel\inst{\ref{inst:15}} $^{\href{https://orcid.org/0000-0003-2144-4316}{\includegraphics[scale=0.5]{figures/orcid.jpg}}}$\and
E. Villaver\inst{\ref{inst:16}}\and
I. Walter\inst{\ref{inst:54}} $^{\href{https://orcid.org/0000-0002-5839-1521}{\includegraphics[scale=0.5]{figures/orcid.jpg}}}$\and 
N. A. Walton\inst{\ref{inst:55}} $^{\href{https://orcid.org/0000-0003-3983-8778}{\includegraphics[scale=0.5]{figures/orcid.jpg}}}$}

\institute{
\label{inst:1} Instituto de Astrofisica e Ciencias do Espaco, Universidade do Porto, CAUP, Rua das Estrelas, 4150-762 Porto, Portugal \and
\label{inst:2} Departamento de Fisica e Astronomia, Faculdade de Ciencias, Universidade do Porto, Rua do Campo Alegre, 4169-007 Porto, Portugal \and
\label{inst:3} Dipartimento di Fisica, Universita degli Studi di Torino, via Pietro Giuria 1, I-10125, Torino, Italy \and
\label{inst:4} Physikalisches Institut, University of Bern, Gesellschaftsstrasse 6, 3012 Bern, Switzerland \and
\label{inst:5} Center for Space and Habitability, University of Bern, Gesellschaftsstrasse 6, 3012 Bern, Switzerland \and
\label{inst:6} Department of Physics and Kavli Institute for Astrophysics and Space Research, Massachusetts Institute of Technology, Cambridge, MA 02139, USA \and
\label{inst:7} Observatoire Astronomique de l'Université de Genève, Chemin Pegasi 51, CH-1290 Versoix, Switzerland \and
\label{inst:8} Department of Astronomy, Stockholm University, AlbaNova University Center, 10691 Stockholm, Sweden \and
\label{inst:9} Aix Marseille Univ, CNRS, CNES, LAM, 38 rue Frédéric Joliot-Curie, 13388 Marseille, France \and
\label{inst:10} Division Technique INSU, CS20330, 83507 La Seyne sur Mer cedex, France \and
\label{inst:11} Centre for Exoplanet Science, SUPA School of Physics and Astronomy, University of St Andrews, North Haugh, St Andrews KY16 9SS, UK \and
\label{inst:12} Space Research Institute, Austrian Academy of Sciences, Schmiedlstrasse 6, A-8042 Graz, Austria \and
\label{inst:13} Cavendish Laboratory, JJ Thomson Avenue, Cambridge CB3 0HE, UK \and
\label{inst:14} Astrobiology Research Unit, Université de Liège, Allée du 6 Août 19C, B-4000 Liège, Belgium \and
\label{inst:15} Space sciences, Technologies and Astrophysics Research (STAR) Institute, Université de Liège, Allée du 6 Août 19C, 4000 Liège, Belgium \and
\label{inst:24} Observatoire astronomique de l'Université de Genève, Chemin Pegasi 51, 1290 Versoix, Switzerland \and
\label{inst:16} Instituto de Astrofisica de Canarias, Via Lactea s/n, 38200 La Laguna, Tenerife, Spain \and
\label{inst:17} Departamento de Astrofisica, Universidad de La Laguna, Astrofísico Francisco Sanchez s/n, 38206 La Laguna, Tenerife, Spain \and
\label{inst:18} Institut de Ciencies de l'Espai (ICE, CSIC), Campus UAB, Can Magrans s/n, 08193 Bellaterra, Spain \and
\label{inst:19} Institut d’Estudis Espacials de Catalunya (IEEC), Gran Capità 2-4, 08034 Barcelona, Spain \and
\label{inst:20} European Space Agency (ESA), European Space Research and Technology Centre (ESTEC), Keplerlaan 1, 2201 AZ Noordwijk, The Netherlands \and
\label{inst:21} Admatis, 5. Kandó Kálmán Street, 3534 Miskolc, Hungary \and
\label{inst:22} Depto. de Astrofisica, Centro de Astrobiologia (CSIC-INTA), ESAC campus, 28692 Villanueva de la Cañada (Madrid), Spain \and
\label{inst:23} Sub-department of Astrophysics, Department of Physics, University of Oxford, Oxford, OX1 3RH, UK \and
\label{inst:25} Max Planck Institut für Extraterrestrische Physik, Gießenbachstraße 1, 85748 Garching bei München, Germany \and
\label{inst:26} INAF, Osservatorio Astronomico di Padova, Vicolo dell'Osservatorio 5, 35122 Padova, Italy \and
\label{inst:27} Université Grenoble Alpes, CNRS, IPAG, 38000 Grenoble, France \and
\label{inst:28} Université de Paris Cité, Institut de physique du globe de Paris, CNRS, 1 Rue Jussieu, F-75005 Paris, France \and
\label{inst:29} Institute of Planetary Research, German Aerospace Center (DLR), Rutherfordstrasse 2, 12489 Berlin, Germany \and
\label{inst:30} INAF, Osservatorio Astrofisico di Torino, Via Osservatorio, 20, I-10025 Pino Torinese To, Italy \and
\label{inst:31} Centre for Mathematical Sciences, Lund University, Box 118, 221 00 Lund, Sweden \and
\label{inst:33} Centre Vie dans l’Univers, Faculté des sciences, Université de Genève, Quai Ernest-Ansermet 30, 1211 Genève 4, Switzerland \and
\label{inst:34} Thüringer Landessternwarte Tautenburg, Sternwarte 5, D-07778 Tautenburg, Germany \and
\label{inst:35} Leiden Observatory, University of Leiden, PO Box 9513, 2300 RA Leiden, The Netherlands \and
\label{inst:36} Department of Space, Earth and Environment, Chalmers University of Technology, Onsala Space Observatory, 439 92 Onsala, Sweden \and
\label{inst:37} Department of Astrophysics, University of Vienna, Türkenschanzstrasse 17, 1180 Vienna, Austria \and
\label{inst:38} Konkoly Observatory, Research Centre for Astronomy and Earth Sciences, 1121 Budapest, Konkoly Thege Miklós út 15-17, Hungary \and
\label{inst:39} ELTE E\"otv\"os Lor\'and University, Institute of Physics, P\'azm\'any P\'eter s\'et\'any 1/A, 1117 Budapest, Hungary \and
\label{inst:40} IMCCE, UMR8028 CNRS, Observatoire de Paris, PSL Univ., Sorbonne Univ., 77 av. Denfert-Rochereau, 75014 Paris, France \and
\label{inst:41} Institut d'astrophysique de Paris, UMR7095 CNRS, Université Pierre \& Marie Curie, 98bis blvd. Arago, 75014 Paris, France \and
\label{inst:42} Astrophysics Group, Lennard Jones Building, Keele University, Staffordshire, ST5 5BG, United Kingdom \and
\label{eso} European Southern Observatory, Karl-Schwarzschild-Straße 2, Garching bei München D-85748, Germany \and
\label{inst:43} Mullard Space Science Laboratory, University College London, Holmbury St. Mary, Dorking, Surrey, RH5 6NT, UK \and
\label{inst:44} INAF, Osservatorio Astrofisico di Catania, Via S. Sofia 78, 95123 Catania, Italy \and
\label{inst:45} Institute of Optical Sensor Systems, German Aerospace Center (DLR), Rutherfordstrasse 2, 12489 Berlin, Germany \and
\label{inst:46} Dipartimento di Fisica e Astronomia "Galileo Galilei", Universita degli Studi di Padova, Vicolo dell'Osservatorio 3, 35122 Padova, Italy \and
\label{inst:47} Department of Physics, University of Warwick, Gibbet Hill Road, Coventry CV4 7AL, United Kingdom \and
\label{inst:48} ETH Zurich, Department of Physics, Wolfgang-Pauli-Strasse 2, CH-8093 Zurich, Switzerland \and
\label{inst:49} Zentrum für Astronomie und Astrophysik, Technische Universität Berlin, Hardenbergstr. 36, D-10623 Berlin, Germany \and
\label{inst:50} Institut fuer Geologische Wissenschaften, Freie Universitaet Berlin, Maltheserstrasse 74-100,12249 Berlin, Germany \and
\label{inst:51} Université de Liège, Allée du 6 Août 19C, 4000 Liège, Belgium \and
\label{inst:52} ELTE E\"otv\"os Lor\'and University, Gothard Astrophysical Observatory, 9700 Szombathely, Szent Imre h. u. 112, Hungary \and
\label{inst:53} HUN-REN--ELTE Exoplanet Research Group, 9700 Szombathely, Szent Imre h. u. 112, Hungary \and
\label{inst:54} German Aerospace Center (DLR), Institute of Optical Sensor Systems, Rutherfordstraße 2, 12489 Berlin \and
\label{inst:55} Institute of Astronomy, University of Cambridge, Madingley Road, Cambridge, CB3 0HA, United Kingdom
}

    \date{}
   
 
  \abstract
   {The HD 15337 (TIC 120896927, TOI-402) system was observed by the Transiting Exoplanet Survey Satellite (TESS), revealing the presence of two short-period planets situated on opposite sides of the radius gap. This offers an excellent opportunity to study formation and evolution theories, as well as investigate internal composition and atmospheric evaporation.}
   {We aim to constrain the internal structure and composition of HD 15337 b and c -- two short-period planets situated on opposite sides of the radius valley -- using new transit photometry and radial velocity data.}
   {We acquire 6 new transit visits with the CHaracterising ExOPlanet Satellite (CHEOPS) and 32 new radial velocity measurements from the High Accuracy Radial Velocity Planet Searcher (HARPS) to improve the accuracy of the mass and radius estimates for both planets. We reanalyse light curves from TESS sectors 3 and 4 and analyse new data from sector 30, correcting for long-term stellar activity. Subsequently, we perform a joint fit of the TESS and CHEOPS light curves, and all available RV data from HARPS and the Planet Finder Spectrograph (PFS). Our model fits the planetary signals, the stellar activity signal and the instrumental decorrelation model for the CHEOPS data simultaneously. The stellar activity was modelled using a Gaussian-process regression on both the RV and activity indicators. We finally employ a Bayesian retrieval code to determine the internal composition and structure of the planets.}
   {We derive updated and highly precise parameters for the HD 15337 system. Our improved precision on the planetary parameters makes HD 15337 b one of the most precisely characterised rocky exoplanets, with radius and mass measurements achieving a precision better than 2\% and 7\%, respectively. We are able to improve the precision of the radius measurement of HD 15337 c to 3\%. Our results imply that the composition of HD 15337 b is predominantly rocky, while HD 15337 c exhibits a gas envelope with a mass of at least $0.01\ M_\oplus$.}
   {Our results lay the groundwork for future studies, which can further unravel the atmospheric evolution of these exoplanets and give new insights into their composition and formation history and the causes behind the radius gap.}
   

    \maketitle
%


\section{Introduction}

The search for exoplanets orbiting solar-type stars has led to a large increase in the number of known planets in recent years. Space missions like Kepler \citep{borucki2016kepler} and the Transiting Exoplanet Survey Satellite \citep[TESS;][]{ricker2014transiting} have been directly responsible for this increase, leading to the detection of many multi-planet systems that are composed of planets in the super-Earth ($R_\mathrm{p} = 1-2\ R_\oplus$) and sub-Neptune ($R_\mathrm{p} = 2-4\ R_\oplus$) regimes. Multi-planet systems that contain two or more small planets with similar masses are of particular importance to understand the formation and evolution of exoplanetary systems and even our own Solar System, by studying the differences in structure and composition between each planet. In some systems (e.g. HD 3167, \citeauthor{gandolfi2017transiting} \citeyear{gandolfi2017transiting}; HD 23472, \citeauthor{barros2022hd} \citeyear{barros2022hd}), the planets lie on opposite sides of the radius gap \citep{fulton2017california}, a gap in the radius distribution in which not many planets have been detected. Some possible explanations for this apparent gap include the effect of atmospheric photoevaporation \citep[e.g.][]{owen2017evaporation, venturini2020nature}, gas-poor formation \citep[e.g.][]{lee2016breeding}, or giant impact erosion \citep[e.g.][]{liu2015giant}. Atmospheric evaporation is the most accepted explanation for this gap. Since the presence of a gas envelope is enough to significantly change the radius of a planet while maintaining a similar mass, it is possible that the smaller planets lost their envelopes due to stellar irradiation. However, \cite{luque2022density} suggest the gap is not a radius gap but a density gap, and that orbital migration of water-rich worlds can explain the observations and the population of sub-Neptunes.\par

The CHaracterising ExOPlanet Satellite \citep[CHEOPS;][]{benz2021cheops} is an ESA space telescope designed as a follow-up mission aiming at the precise characterisation of exoplanetary systems around nearby bright stars. It was launched on December 18, 2019, and has been orbiting at $\sim$700 km above Earth since then. The increased precision in the photometry measurements makes CHEOPS observations a valuable addition to previously obtained transiting light curves (from TESS or ground-based observatories), which leads to a highly precise radius and an improvement in the internal characterisation of the planets.\par
HD 15337 (TIC 120896927, TOI-402) is a bright (V=9) K1 dwarf, known to host two planets lying on opposite sides of the radius valley \citep{gandolfi2019,dumusque2019hot}, which were first detected using TESS observations and confirmed with HARPS radial velocity (RV) measurements. HD 15337 b is a short-period ($P=4.76$ d) super-Earth with $R=1.78R_\oplus$ and $M=6.5\ M_\oplus$ with a companion sub-Neptune (HD 15337 c) with a 17.2-day period and a similar mass ($M=6.7\ M_\oplus$) but a larger radius ($R=2.5 R_\oplus$), thought to have a gaseous envelope. As such, it is one of the most amenable systems to study the physics behind the radius valley due to the characteristics of both planets and the brightness of the star.\par
In this paper, we use archive and new TESS data and newly obtained CHEOPS photometric observations, together with archive and new HARPS and PFS ground-based RV measurements, to improve the precision of the radius and the mass of HD 15337 b and HD 15337 c. We present a summary of the observations and data reduction methods in Section 2, followed by the estimation of the stellar parameters in Section 3. In Section 4, we describe our models and assumptions and summarise the new results. We discuss our results in Section 5 and finalise with a short conclusion in Section 6. 


\section{Observations}

\subsection{CHEOPS}

We obtained 6 CHEOPS visits of HD 15337 as part of the CHEOPS Guaranteed Time Observation (GTO) programme, for a total observation time of $\sim$2.2 d. We obtained 3 transits of planet b and 2 transits of planet c, plus an overlapping transit of both planets during the first visit. The details of each visit are summarised in the observation log in Table \ref{log}.\par
The data of each visit were reduced with the CHEOPS data reduction pipeline\footnote{DRP version 13} \citep[DRP;][]{hoyer2020expected}, which processes all the data automatically and corrects for bias, gain, dark, flat, and environmental effects such as cosmic ray impacts, background, and smearing. The DRP extracts the photometric signal in four apertures, RINF, DEFAULT, RSUP and OPTIMAL, which calculates the optimal radius based on maximising the signal-to-noise ratio of the light curve. It then produces a file with all the extracted light curves and additional data such as the roll angle time series, quality flags and the background time series used for the corrections. We chose the DEFAULT setting, with a radius of 25pix, that was considered to be the best according to the data reduction report.

\begin{table*}
\caption{CHEOPS observation log for HD 15337.}
\label{log}      
\centering
\resizebox{\textwidth}{!}{
\begin{tabular}{c c c c c c c c}       
\hline\hline
ID & Planets & File Key & Start Date [UTC] & Duration [h] & Exp. Time [s] & Efficiency & No. Points\\
1 & b, c & CH\_PR100031\_TG044201\_V0200 & 2021-09-04T21:41:56 & 8.57 & 38.6 & 66.2\% & 530\\
2 & b & CH\_PR100031\_TG045301\_V0200 & 2021-10-22T08:05:36 & 10.76 & 38.6 & 91.5\% & 919\\
3 & c & CH\_PR100031\_TG045401\_V0200 & 2021-10-26T09:56:35 & 7.58 & 38.6 & 94.8\% & 671\\
4 & b & CH\_PR100031\_TG045302\_V0200 & 2021-11-05T16:50:36 & 9.51 & 38.6 & 83.4\% & 740\\
5 & b & CH\_PR100031\_TG045303\_V0200 & 2021-11-10T09:56:37 & 9.51 & 38.6 & 81.5\% & 723\\
6 & c & CH\_PR100031\_TG045402\_V0200 & 2021-11-12T13:57:57 & 7.69 & 38.6 & 75.6\% & 543\\
\hline
\end{tabular}}
\tablefoot{The first column shows the number of the observation as referred to in this paper. The second column shows the planet transits observed in each visit and the remaining columns show the unique file key of each CHEOPS visit, the start date of the observations, the duration, the exposure time of each observation, the efficiency and the number of non-flagged data points on each visit.}

\end{table*}

\subsection{TESS}

HD 15337 was previously observed by Camera \#2 of TESS with a two-minute cadence in Sector 3 (from 2018-Sep-20 to 2018-Oct-18) and Sector 4 (2018-Oct-18 to 2018-Nov-15), as reported in \cite{gandolfi2019} and \cite{dumusque2019hot}, and more recently in Sector 30 (2020-Sep-22 to 2020-Oct-21) which we include in this analysis. A total of 13 transits of HD 15337 b and five of HD 15337 c were found in the three sectors.\\

The data were reduced by the Science Processing Operations Center \citep[SPOC;][]{jenkins2020kepler} pipeline and the fits files were downloaded from the Mikulski Archive for Space Telescopes (MAST) portal\footnote{\url{https://mast.stsci.edu/}.}. In this analysis, we employ the Presearch Data Conditioning Simple Aperture Photometry (PDCSAP) light curves, which are corrected to remove instrumental systematics, outliers, discontinuities, and other non-astrophysical long-term trends from the data \citep{smith2012kepler, smith2017kepler}.\par
The files obtained from the SPOC pipeline contain a quality parameter that identifies data points that have been impacted by abnormal occurrences including attitude changes, momentum dumps and thruster firings \citep{tenenbaum2018tess}. To ensure reliable data, we deleted all the bad-quality flagged data points from our light curve as well as any NaN values present before detrending and processing the data.
We followed a similar approach to the one described in \cite{rosario2022measuring} for the correction of the TESS light curves. We isolated each individual transit, keeping all points within three transit durations of the estimated mid-transit time and removing the remaining out-of-transit data. We normalised each transit light curve separately to avoid the influence of long-term trends, using a low-order polynomial fit. We used the Bayesian Inference Criterion (BIC) to choose the order of the polynomial for each transit, with a linear trend being preferred in the majority of the cases.

\subsection{HARPS}

HD 15337 was Doppler-monitored between 15 December 2003 and 6 September 2017 UT using the High Accuracy Radial Velocity Planet Searcher (HARPS) spectrograph \citep[$R\approx115\,000$;][]{Mayor2003} mounted at the ESO-3.6\,m telescope \citep[La Silla Observatory; see][]{gandolfi2019, dumusque2019hot}. We retrieved the 87 publicly available data from the ESO archive and acquired 32 additional HARPS spectra between 9 July and 12 September 2019 UT, as part of our follow-up program of TESS transiting planets (ID: 1102.C-0923; PI: Gandolfi). We set the exposure time to $T_\mathrm{exp}$\,=\,900-1800\,s based on seeing and sky conditions, leading to a median signal-to-noise (S/N) ratio of $\sim$92 per pixel at 550\,nm. We reduced the archival and new HARPS data using the dedicated data reduction software \citep[DRS; ][]{Lovis2007} and extracted the radial velocity (RV) measurements cross-correlating the Echelle spectra with a K5 numerical mask \citep{Baranne1996,Pepe2002}. We also used the DRS to extract the full-width half maximum (FWHM) and bisector inverse slope (BIS) of the cross-correlation function (CCF) and the Ca\,{\sc ii} H\,\&\,K lines activity index\footnote{We adopted a B$-$V colour index of 0.880.} (log\,$R^\prime_\mathrm{HK}$). In June 2015, the HARPS spectrograph was upgraded by replacing the circular fibres with octagonal ones \citep{LoCurto2015}. To account for the RV offset due to the instrument refurbishment, we treated the archive HARPS RV measurements taken before (52 measurements) and after (67 measurements) June 2015 as independent data sets.\par

The HARPS RVs were corrected for secular acceleration following the equations from \cite{kuerster2003low}. We retrieved the stellar RV and proper motion from Gaia DR2 \citep{soubiran2018gaia} and EDR3 \citep{collaboration2020vizier} respectively and found the average secular acceleraton during the HARPS observations timeframe to be 0.05189 ms$^{-1}$yr$^{-1}$. Given the 15.743 yr baseline of the HARPS data, the secular acceleration implies a correction up to about +0.82 ms$^{-1}$ for the last data point in our time-series.

\subsection{PFS}

HD 15337 was observed as part of the Magellan-TESS Survey (MTS) published in \cite{teske2021magellan}. The MTS followed up several pre-selected planets from TESS to obtain RVs using the Planet Finder Spectrograph \citep[PFS;][]{crane2006carnegie, crane2008carnegie, crane2010carnegie} located on the Magellan II Clay telescope at the Las Campanas Observatory. PFS is a precision RV spectrograph commissioned in October 2009, used primarily for the search of new planets and follow-up transit planet candidates. The PFS detector, after January 2018, has a 10k $\times$ 10k CCD detector with 9$\mu$m pixels and a resolving power of $\approx$130,000. HD 15337 was observed between 12 February 2019 and 20 December 2019 and 48 new RV measurements were obtained. The spectra were reduced in \cite{teske2021magellan} using a custom pipeline based on the one from \cite{butler1996attaining}. The PFS RVs were included as a separate dataset in our analysis to account for the different offset.


\section{Stellar characterisation}

The spectroscopic stellar parameters for HD 15337 (effective temperature $T_{\mathrm{eff}}$, surface gravity $\log g$, microturbulence velocity, iron content [Fe/H]) were taken from a previous version of SWEET-Cat \citep[][]{Santos-13, sousa2018sweet}. The parameters were estimated based on a combined HARPS spectrum with the ARES+MOOG methodology using the latest version of ARES\footnote{\textit{A}utomatic \textit{R}outine for Line \textit{E}quivalent Widths in Stellar \textit{S}pectra. The latest version, ARES v2, can be downloaded at \url{https://github.com/sousasag/ARES}.} \citep{Sousa-07, Sousa-15} to consistently measure the equivalent widths (EW) of selected iron lines on the spectrum. In this analysis, we use a minimisation process to find the ionisation and excitation equilibrium to converge for the best set of spectroscopic parameters. This process makes use of a grid of Kurucz model atmospheres \citep{Kurucz-93} and the radiative transfer code MOOG \citep{Sneden-73}. Since the star is cooler than 5200~K we used the appropriate iron line list for our method presented in \citet[][]{Tsantaki-13}. More recently the same methodology was applied on a more recent combined HARPS spectrum where we derived new spectroscopic stellar parameters ($T_{\mathrm{eff}} = 5088 \pm 78$ K, $\log g = 4.24 \pm 0.10$ (dex), and [Fe/H] $0.04 \pm 0.03$ dex; \citep[][]{Sousa-21}), consistent within 2.2$\sigma$. Here, we also derived a more accurate trigonometric surface gravity using recent GAIA DR3 data \citep{gaia2016gaia, gaia2023gaia} following the same procedure as described in \citet[][]{Sousa-21}, which provided a consistent value when compared with the spectroscopic surface gravity (4.48 $\pm$ 0.04 dex). Abundances of magnesium (Mg) and silicon (Si) were also derived using the same tools and models as for the stellar parameter determination, as well as using the classical curve-of-growth analysis method assuming local thermodynamic equilibrium. For the derivation of abundances, we closely followed the methods described in \citet[][]{Adibekyan-12, Adibekyan-15}.

We determined the radius of HD 15337 using a Markov-Chain Monte Carlo (MCMC) modified infrared flux method \citep{Blackwell1977,Schanche2020}. This was done by building spectral energy distributions (SEDs) from stellar atmospheric models defined by our spectral analysis and calculating the stellar bolometric flux by comparing synthetic and observed broadband photometry in the following bandpasses: {\it Gaia} $G$, $G_\mathrm{BP}$, and $G_\mathrm{RP}$, 2MASS $J$, $H$, and $K$, and \textit{WISE} $W1$ and $W2$ \citep{Skrutskie2006,Wright2010,GaiaCollaboration2021}. Using known physical relations we therefore obtained the stellar effective temperature and angular diameter that is converted to a radius using the offset-corrected \textit{Gaia} parallax \cite{Lindegren2021}. To correctly estimate the error in our stellar radius we conducted a Bayesian modelling averaging of the \textsc{atlas} \citep{Kurucz-93,Castelli2003} and \textsc{phoenix} \citep{Allard2014} catalogues to produce a weighted averaged posterior distribution of the radius that encapsulates uncertainties in stellar atmospheric modelling. From this analysis we obtained $R_s=0.855\pm0.008\,R_{\odot}$.

We then derived the isochronal mass $M_s$ and age $t_s$ by inputting $T_{\mathrm{eff}}$, [Fe/H], and $R_s$ along with their uncertainties into two different stellar evolutionary models. In detail, we computed a first pair of mass and age estimates ($M_{s,1}\pm\sigma_{M1}$, $t_{s,1}\pm\sigma_{t1}$) through the isochrone placement algorithm \citep{bonfanti15,bonfanti16}, which interpolates the input set within pre-computed grids of PARSEC\footnote{\textsl{PA}dova and T\textsl{R}ieste \textsl{S}tellar \textsl{E}volutionary \textsl{C}ode: \url{http://stev.oapd.inaf.it/cgi-bin/cmd}.} v1.2S \citep{marigo17} tracks and isochrones. Further providing $P_{\mathrm{rot}}=36.5$ d \citep{gandolfi2019} to our interpolating routine, we coupled the isochrone fitting with the gyrochronological relation from \citet{barnes2010}, which is implemented in our isochrone placement to improve the optimisation scheme convergence as detailed in \citet{bonfanti16} and retrieve more precise stellar parameters \citep[see e.g.][]{angus2019}. A second pair ($M_{s,2}\pm\sigma_{M2}$, $t_{s,2}\pm\sigma_{t2}$) was derived by the CLES \citep[Code Liègeois d'Évolution Stellaire][]{scuflaire08} code, which generates the best-fit evolutionary track according to the input set of stellar parameters following the Levenberg-Marquadt minimisation scheme \citep[see e.g.][]{salmon21}. We successfully checked the mutual consistency of the two respective pairs of outcomes via the $\chi^2$-based criterion described in \citet{bonfanti21} and then we merged the two probability density functions inferred from both the two pairs ($M_{s,1}\pm\sigma_{M1}$, $M_{s,2}\pm\sigma_{M2}$) and ($t_{s,1}\pm\sigma_{t1}$, $t_{s,2}\pm\sigma_{t2}$) to finally obtain $M_s=0.840\pm0.041\,M_{\odot}$ and $t_s=9.6_{-3.9}^{+3.8}$ Gyr. See \citet{bonfanti21} for further details about the statistical methodology we followed.

Finally, we note that HD 15337 has a very dim stellar companion \citep[$\Delta m \sim 9.33$ mag at 832 nm,][]{Lester21}, detected via speckle imaging at an angular separation of 1.4'' from the primary \citep{Ziegler20, Lester21}. The inferred stellar mass ratio (mass of the secondary over the mass of the primary) is 0.14 \citep{Lester21}. The presence of a blended stellar companion causes a dilution factor in the planetary radii, which, in case the planets orbit the primary star, can be computed using eq. (7) of \citet{Ciardi15}. Using that equation, we find a correction factor of 1.00009 (as long as the planets orbit the primary). 
For this particular system, it is rather straightforward to conclude that the planets orbit the primary. If one of the planets was orbiting the secondary, the maximum possible transit depth would correspond to a configuration where such planet completely blocks the flux stemming from the secondary star. This would yield a maximum transit depth in our light curves of $\delta_{\rm max}\sim 185$ ppm\footnote{Value inferred using the definition of the transit depth for a binary system (i.e, with the two stellar fluxes in the denominator), and imposing the radius of the planet equal to the radius of the secondary, as an extreme limiting case.}. The transit depth that we measure with CHEOPS for planet b is $\delta \approx 364$ ppm and $\delta\approx 735$ ppm for planet c. Thus, the transits are too deep to correspond to the secondary star and we can confidently conclude that the two planets orbit the primary. The correction factor for the planetary radii is sufficiently small for our interior characterisation analysis not to be affected by the small dilution effect caused by the stellar companion.

\begin{table}
\centering
\caption{An overview of fundamental stellar parameters for HD 15337.}
\label{stepartable}
\begin{tabular}{lll}
\hline
Parameter [unit] & Value & Source\\
\hline
Name                                  & HD 15337                       & --    \\
\textit{Gaia} DR3 ID                  & 5068777809824976256           & G2022 \\
$G$ (\textit{Gaia}) [mag]             & $8.865 \pm 0.00276$          & G2022 \\
$T_\mathrm{eff}$ [K]                  & $5131  \pm 74$                & This work\\
$\log g$ [cgs]                        & $4.37  \pm 0.08$              & This work (spec)\\
$\log g$ [cgs]                        & $4.48  \pm 0.04$              & This work (Gaia)\\
$[\mathrm{Fe/H}]$ [dex]               & $0.03  \pm 0.04$              & This work\\
$[\mathrm{Mg/H}]$ [dex]               & $0.09  \pm 0.06$              & This work\\
$[\mathrm{Si/H}]$ [dex]               & $0.07  \pm 0.05$              & This work\\
$v_\mathrm{mic}$ [km~s$^{-1}$]        & $0.87  \pm 0.13$              & This work\\ 
$R_\mathrm{s}$ [$R_\odot$]            & $0.855 \pm 0.008$             & This work\\
$M_\mathrm{s}$ [$M_\odot$]            & $0.840 \pm 0.041$             & This work\\
$t_\mathrm{s}$ [Gyr]                  & $9.6_{-3.9}^{+3.8}$          & This work\\
\hline
\end{tabular}
\tablefoot{G2022 = \citet{GaiaCollaboration2022}}

\end{table}


\section{Data analysis}

\subsection{Detrending of CHEOPS light curves}\label{4_1_decorrelation}

\begin{figure}
     \centering
        \includegraphics[width=\linewidth]{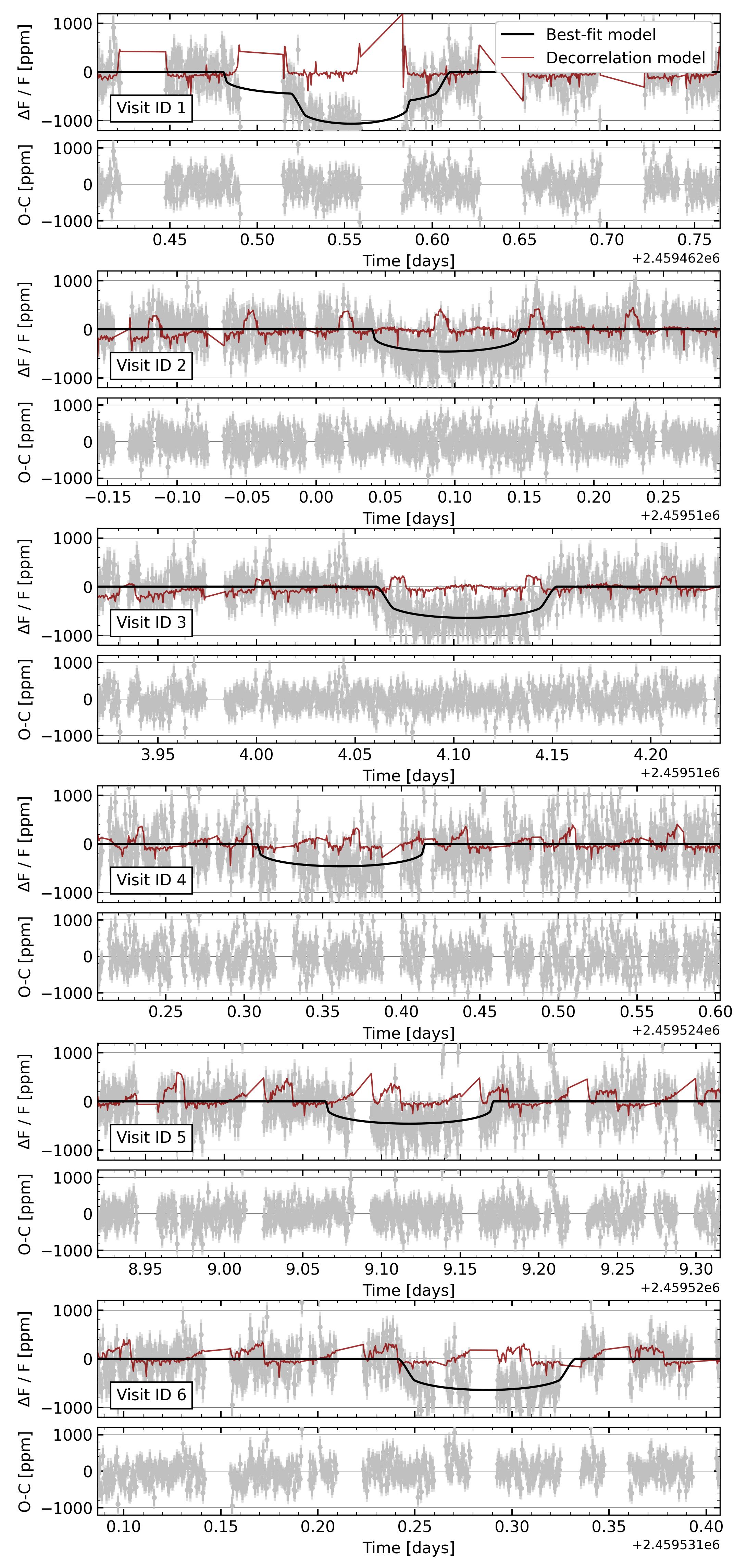}
        \caption{CHEOPS light curves with spline decorrelation model (red) and best-fit transit model (black). The data from each CHEOPS visit (after removing outliers) is plotted in grey. Visit 1 has an overlapping transit of planets b and c, visits 2, 4 and 5 capture a transit of planet b and visits 3 and 6 capture planet c.}
        \label{cheopslcs}
\end{figure}

The light curves shown in Figure \ref{cheopslcs} show strong correlations with the roll angle as well as other systematic effects. CHEOPS is located in a nadir-locked low-Earth orbit, which means the field stars rotate around the target star as a function of the position of the spacecraft in its orbit. This also leads to a correlation between the background flux and the roll angle, especially when close to an Earth occultation. The gaps in the observations can be caused by Earth occultations due to the CHEOPS orbit placement or by South Atlantic Anomaly (SAA) crossings.\par
Another relevant effect for the majority of the targets observed by CHEOPS is temperature variation. Temperature perturbations are caused mainly by Earth occultations or when moving between targets due to the position of CHEOPS relative to the sun. This causes a ramp effect at the beginning of each time series that is discussed in more detail in \cite{morris2021cheops} and is corrected by detrending against the temperature of the telescope tube using the ThermFront2 Sensor temperature parameter included in the CHEOPS light curves.\par
We first removed outliers by performing a sigma-clip at 3$\sigma$ using the centroid X and Y time series and at 6$\sigma$ in the flux, ensuring we did not remove any part of the transit. The centroid position can be affected by cosmic rays, stray light or satellite trails crossing the point spread function, which helps to identify outliers. Points that varied more than 10\% from the background-flux time series were also removed.\par
To mitigate the impact of these instrumental variations on the light curve, we perform a spline decorrelation simultaneously to the transit model fitting with the code LISA \citep{demangeon2018discovery,demangeon2021warm}. The spline decorrelation consists of a sequential fit of the residuals against the time series of the roll angle, centroid position, temperature of the telescope tube and measured background.\par

\subsection{Radial velocity periodogram analysis}

\begin{figure*}
     \centering
        \includegraphics[width=\linewidth]{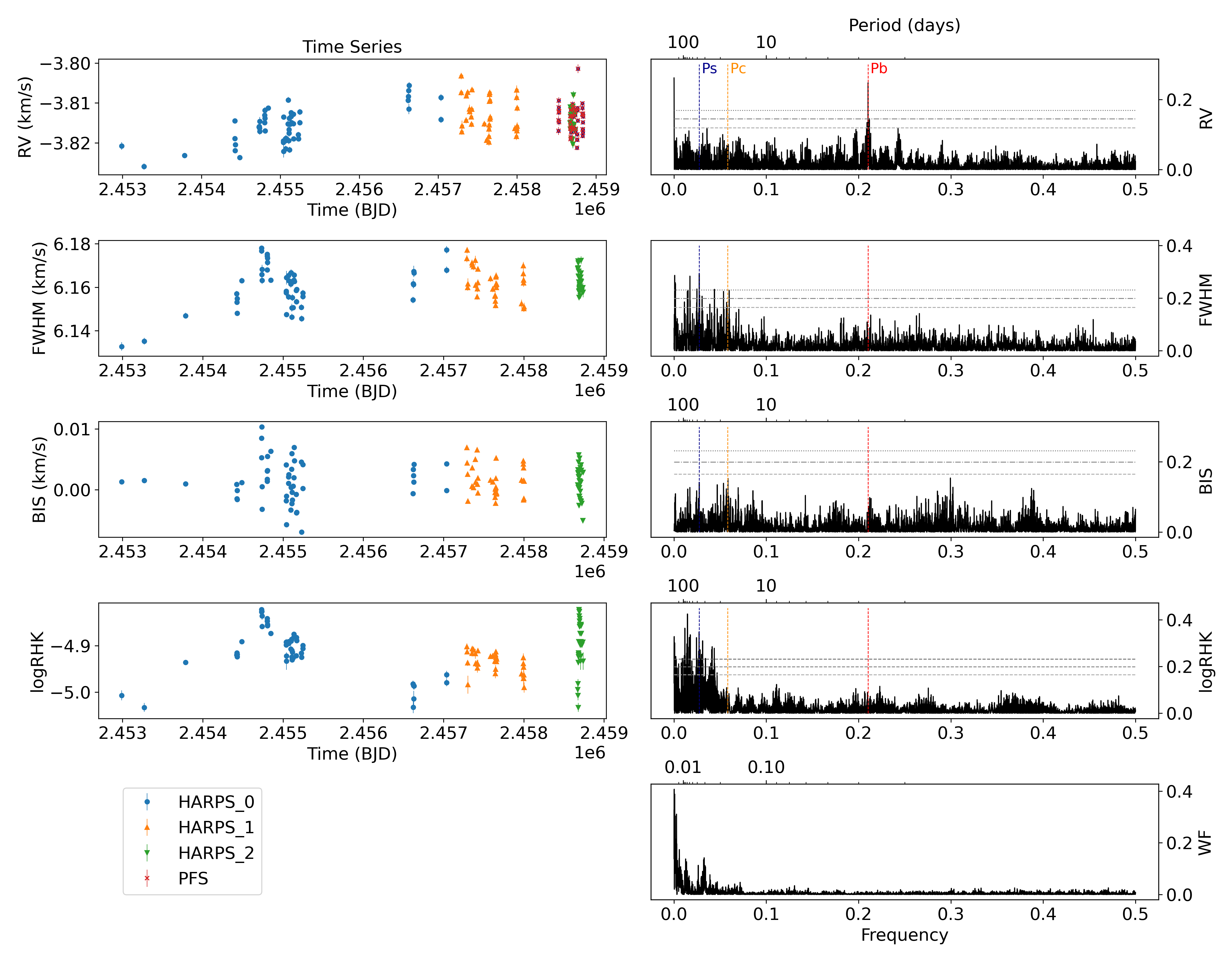}
        \caption{\textit{Left}: Plot of the offset-corrected time series of the RV observations of HARPS and PFS, as well as the activity indicators of the HARPS RVs. For a preliminary analysis and plotting purposes, the offset between the different RV sets was estimated as the median of each time series. \textit{Right}: Corresponding GLS periodograms, computed from the offset-corrected values, as the offsets induce additional non-physical peaks that can potentially mask other relevant peaks. The window function is plotted in the last row. The vertical dashed lines show the approximate periods of the known planets as well as the expected stellar rotation period. The dashed horizontal lines show the 10\% (dashed line), 1\% (dot-dashed line), and 0.1\% (dotted line) false-alarm probabilities as per \cite{zechmeister2009generalised}.}
        \label{period_init}
\end{figure*}

\begin{figure}
     \centering
        \includegraphics[width=\linewidth]{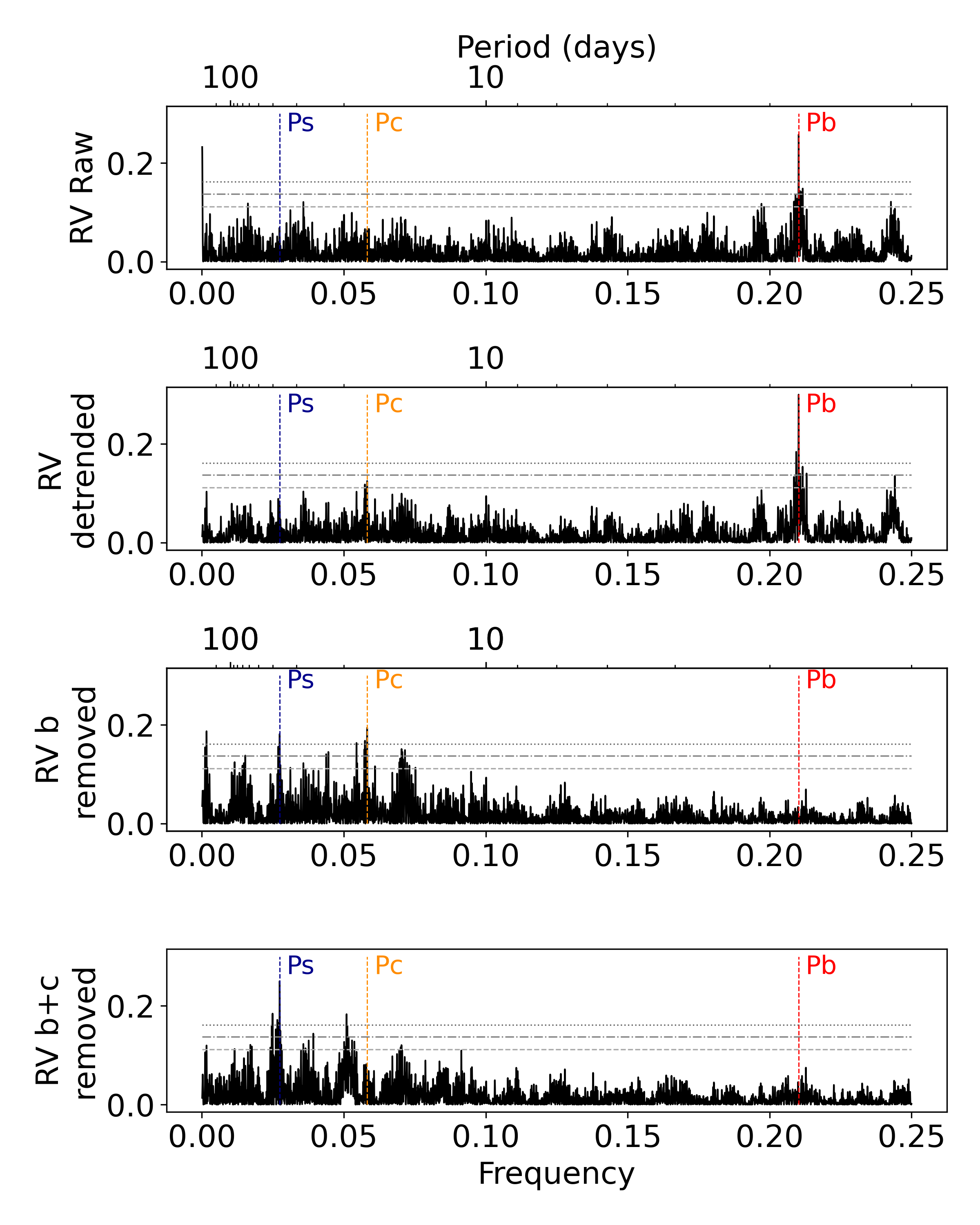}
        \caption{GLS periodograms, computed from the offset-corrected values. The upper panel shows the raw periodogram of the RVs as shown in Figure \ref{period_init}. The second panel shows the periodogram after the removal of the long-term trends (removed with a simple second-degree polynomial regression for pre-whitening purposes only). The lower panels show the periodograms after the removal of the Keplerian signals of HD 15337 b and c from the previous one. Vertical and horizontal lines as Figure \ref{period_init}.}
        \label{period_planets}
\end{figure}

Figure \ref{period_init} shows the offset-corrected radial velocity (RV) time series from HARPS and PFS, as well as several activity indicators obtained from the HARPS data reduction: FWHM, BIS and $\log{R'_\mathrm{HK}}$. The generalised Lomb Scargle \citep[GLS;][]{zechmeister2009generalised} periodogram of each of the aforementioned time series is plotted alongside. A long-term trend is visible in the plotted RVs, which could be caused by long-term stellar activity like a magnetic cycle. The presence of a similar trend in the FWHM time series further hints towards a signal induced by activity. The Pearson correlation factors obtained between the RV and the FWHM series show a strong linear correlation (p-value of $5\times10^{-7}$). However, there appears to be no long-term linear correlation (p-value of 0.25) with the $\log{R'_\mathrm{HK}}$ series.
The main peak in the RV time series clearly shows the presence of a planetary signal around the period of HD 15337 b reported by \cite{gandolfi2019} and \cite{dumusque2019hot}, which is not present in any of the indicators. We do not see, at first glance, an indication of any significant peak near the period of HD 15337 c. To verify its presence, we removed the Doppler signal of the planet by modelling it as a Keplerian with the \texttt{radvel} package, fixing period and phase at the transit ephemerides. Figure \ref{period_planets} shows the periodogram of the RV residuals after removing planet b, in the middle-lower panel, and both planets b and c, in the lower panel. For this preliminary analysis we removed the long-term trends that are shown in the first plot of Figure \ref{period_init} by fitting a simple second-degree polynomial. We see that removing the trends and the signal of planet b, two new peaks are shown, corresponding to the period of planet c and the expected stellar rotation period of 36.5 days \citep{gandolfi2019,dumusque2019hot}. The latter is shown more clearly after removing the signal of planet c in the same way as planet b.\par
The GLS periodograms of the indicators show significant peaks near the stellar rotation period of 36.5 days, but also show peaks that are close to twice that value ($\sim73$ days), which could indicate that 36.5 days is half the stellar rotation period. However, K-type stars with a period this high are unlikely, especially when factoring in the temperature of 5131 K. \cite{mcquillan2014rotation} and \cite{santos2021surface} studied the relationship between the stellar mass, the effective temperature and the rotation period and according to those relationships, a period of 36.5 days is more likely. We also do not find a significant periodic signal at 73 days in the RV residuals after removing the planetary signals (Fig. \ref{period_planets}), which further hints towards the shorter rotation period.
We opted to use the $\log{R'_\mathrm{HK}}$ indicator to model the stellar activity and decorrelate the RVs from HARPS, as described in section \ref{4_3_3_rv}, since the FWHM and BIS periodograms show more significant peaks around the orbital period of planet c and can influence the retrieval of that signal from the RV series. The $\log{R'_\mathrm{HK}}$ was also previously used to mitigate stellar activity in this system by \cite{dumusque2019hot} using part of the same HARPS data we are analysing.

\subsection{Transit and RV joint analysis}

We performed a simultaneous fit of the previously detrended TESS photometry, the raw CHEOPS photometry, the raw HARPS and PFS RVs, the chosen stellar activity indicators and the instrumental decorrelation parameters of CHEOPS (see section \ref{4_1_decorrelation}) using the code LISA \citep{demangeon2018discovery,demangeon2021warm}. The transits are modelled with the \texttt{batman} package \citep{kreidberg2015batman} while the RV modelling uses the \texttt{radvel} package \citep{fulton2018radvel}.  We model the stellar activity using a Gaussian process (GP) as described in section \ref{4_3_3_rv}, which is fitted to the RV and the $\log{R'_\mathrm{HK}}$ data. In this analysis, we use a Bayesian inference framework that maximises the posterior probability. The parameter space is explored with the use of a Markov Chain Monte Carlo (MCMC) algorithm implemented by the package \texttt{emcee} \citep{foreman2013emcee}. We use a number of walkers four times greater than the number of parameters. To avoid a heavy computational load and too much memory usage while saving the chains and walkers, we perform a first initial run with 5000 steps, aimed at obtaining a rough estimation of the best-fit parameters. We then follow up with an additional 10000 steps, starting from the median values from the initial run (excluding burn-in). The prior distributions are described in the following sections and a detailed summary of the values can be found in Table \ref{priors}.

\subsubsection{Transit model} \label{transitmodel}

The transit model for each of the two planets is parameterised by the orbital period (P), the mid-transit time ($T_\mathrm{0}$), the cosine of the inclination ($\cos{i}$), the planet-star radius ratio ($R_\mathrm{p}/R_\star$), the products $e\sin{\omega}$ and $e\cos{\omega}$, where $e$ is the eccentricity and $\omega$ is the argument of the periastron, and the stellar density ($\rho_\star$). The limb-darkening is modelled with a quadratic law, with two coefficients for each instrument ($u_1$ and $u_2$). An additive jitter parameter and an offset parameter for the median flux are included in the model for each TESS sector and each CHEOPS visit.\par
We use a joint prior for the transit parameters ($P$,  $T_\mathrm{0}$, $\cos{i}$, $\rho_*$), described as transiting prior in \cite{demangeon2021warm}. This joint prior effectively uses these parameters to compute the impact parameter ($b$) and the planetary orbital phase ($\phi$) and allows us to define priors on $b$ and $\phi$ instead of $\cos{i}$ and $T_\mathrm{0}$. We set $\phi = 0$ to match the reference time $T_\mathrm{0}$. We set a uniform prior for $P$ and $\phi$ ensuring the chains would not go out of the range of the CHEOPS observations during exploration and use a uniform prior for $R_\mathrm{p}/R_\star$ between 0.01 and 0.03 (for planet b) or 0.04 (for planet c). A uniform distribution between 0 and 2 is picked for $b$ to allow grazing transiting, with the additional condition that $b < 1 + R_\mathrm{p}/R_\star$ to translate our prior knowledge that the planets are transiting. Similar to the transiting prior, a polar joint prior is used to convert $e\sin{\omega}$ and $e\cos{\omega}$ into $e$ and $\omega$. We set a beta distribution as the prior for $e$, with values of $a = 0.867$ and $b = 3.03$ using the formulation from \cite{kipping2013parametrizing}, and a uniform prior between $-\pi$ and $\pi$ for $\omega$. For the limb-darkening coefficients, we set normal prior distributions whose mean and standard deviation are derived with the Limb-Darkening Toolkit \citep[LDTk,][]{parviainen2015ldtk}, which estimates $u_1$ and $u_2$ from the effective temperature ($T_\mathrm{eff}$), gravity ($\log{g}$), and metallicity ([Fe/H]) of the star, taking into account the response function of the instrument. We finally set a uniform prior for the jitter term, between zero and a value five times the median error from each data set, and a normal prior for each offset parameter.

\subsubsection{Radial velocity model} \label{4_3_3_rv}

The model for the radial velocity analysis can be divided into three parts: the planetary model, the stellar activity model and the instrumental detrending. As we know, stellar activity is a major source of uncertainty when looking for planetary signals in RV measurements. We use an approach similar to the one in \cite{demangeon2021warm}, fitting a GP with a quasi-periodic kernel defined by:
\begin{equation}
K(t_i,t_j)=A^2\exp\left[{-\frac{(t_i-t_j)^2}{2\tau_\mathrm{decay}^2}-\frac{\sin^2({\frac{\pi}{P_\mathrm{rot}}|t_i-t_j|})}{2\gamma^2}}\right]
\end{equation}
where $A$ and $P_\mathrm{rot}$ are the amplitude and the period of recurrence of the covariance, $\tau_\mathrm{decay}$ is the decay timescale and $\gamma$ is the periodic coherence scale. The amplitude $A$ is related to the amplitude of the stellar activity signal and $P_\mathrm{rot}$ to the stellar rotation period, making it a key parameter in the stellar activity model. We use two independent Gaussian processes with a quasi-periodic kernel implemented with the \texttt{george} package \citep{2015ITPAM..38..252A} to model the stellar variability: one for the $\log{R'_\mathrm{HK}}$ and the other for the RV data. While independent, the two GPs share some of their hyperparameters. The period of recurrence, the decay timescale and the period coherence scale of the covariance are common to both GPs, while the amplitude of the covariance is different. 
The planetary signals are modelled with Keplerian orbit models and a systemic velocity ($v_\mathrm{0}$). The Keplerian model is parameterised by the semi-amplitude of the RV signal ($K$), $P$, $e\sin{\omega}$ and $e\cos{\omega}$.
For the instrumental model, we define the first set of data (HARPS\_0) as the reference, meaning $v_\mathrm{0}$ is defined by this data set. We then define an offset ($\Delta$RV) between each data set and this reference set, which is particularly important to model the offset caused by the fibre change in HARPS as well as the fact that the PFS RVs are processed to be centred around zero. We fit the same offset value $\Delta RV_\mathrm{HARPS\_1}$ for the two HARPS data sets (67 measurements) after the fibre change. 
We chose to model the long-term trend in the RVs with a long-period sinusoidal function with the period $P_\mathrm{sin}$, the semi-amplitude $K_\mathrm{sin}$ and the reference time $T_{0,\mathrm{sin}}$ set as free parameters. The $\log{R'_\mathrm{HK}}$ time series includes a second-degree polynomial trend with coefficients given by $\log{R'_\mathrm{HK}}_\mathrm{0}$, $\log{R'_\mathrm{HK}}_\mathrm{1}$ and $\log{R'_\mathrm{HK}}_\mathrm{2}$. We also add a jitter parameter as we did for the transit model.

We set a Gaussian prior for each offset parameter, with the mean given by the difference between the average of each data set and a conservative standard deviation of 0.01 km\,s$^{-1}$. As for the transit model, we put a uniform prior in the jitter parameter for each set of RVs, between zero and five times the average of the error of the data sets. The prior for $v_\mathrm{0}$ was defined as a Gaussian distribution centred in the average value of the reference data set (HARPS\_0) with a variance of 0.01 km\,s$^{-1}$ as chosen for the offsets. For the hyperparameters of the kernel used for the Gaussian process model, we define a uniform prior between 0 and 0.01 km\,s$^{-1}$ for the amplitude in the RV kernel and between 0 and 0.1 for the amplitude in the $\log{R'_\mathrm{HK}}$ kernel. The prior for the $P_\mathrm{rot}$ parameter is set as uniform. The decay timescale was set to vary between 10 and 200 days with a uniform prior. For $\gamma$, the typical value is thought to be around 0.5 \citep{dubber2019using}, so we chose a uniform prior between 0.05 and 5, one order of magnitude below and above that value. We also set wide uniform priors for the free parameters of the sinusoidal trend in the RVs. Finally, we chose a uniform prior between 0 and 0.01 km\,s$^{-1}$ for the amplitude of the planetary signals, as seen in \cite{gandolfi2019}. The remaining parameters of the planetary model are shared with the transit model and follow the priors described in section \ref{transitmodel}.

\subsubsection{Joint analysis results}

\begin{table}
\caption{HD 15337 system parameters from the joint fit of transits and RVs.}
\label{tableresults}
\renewcommand{\arraystretch}{1.25}
\begin{tabular}{lc}\hline
Parameters & Derived value\\ \hline
Model stellar parameters\\
$\rho_\star$ ($\rho_\sun$) & $1.334_{-0.079}^{+0.075}$\\
$u_{1,\mathrm{TESS}}$ & $0.4761_{-0.0012}^{+0.0015}$\\
$u_{2,\mathrm{TESS}}$ & $0.1219_{-0.0027}^{+0.0024}$\\
$u_{1,\mathrm{CHEOPS}}$ & $0.6198_{-0.0012}^{+0.0013}$\\
$u_{2,\mathrm{CHEOPS}}$ & $0.0696_{-0.0019}^{+0.0018}$\\
$v_\mathrm{0}$ (km\,s$^{-1}$) & $-3.8292_{-0.0042}^{+0.0033}$\\
\hline

Derived stellar parameters\\
$M_\star$ ($M_\sun$) & $0.829\pm0.038$\\
$R_\star$ ($R_\sun$) & $0.855\pm0.008$\\
$T_\mathrm{eff}$ (K) & $5131\pm74$\\
\hline

Model parameters of HD 15337 b\\
$R_\mathrm{b}/R_\star$ & $0.01898_{-0.00026}^{+0.00030}$\\
$P_\mathrm{b}$ (days) & $4.7559804_{-0.0000037}^{+0.0000062}$\\
$T_\mathrm{0,b}$ (BJD) & $2458411.4623_{-0.0013}^{+0.0008}$\\
$e_\mathrm{b}\cos{\omega_\mathrm{b}}$ & $0.014_{-0.042}^{+0.047}$\\
$e_\mathrm{b}\sin{\omega_\mathrm{b}}$ & $0.043_{-0.014}^{+0.012}$\\
$\cos{i_\mathrm{b}}$ & $0.0123_{-0.0083}^{+0.0100}$\\
$K_\mathrm{b}$ (km\,s$^{-1}$) & $0.00282\pm0.00015$\\
\hline

Derived parameters of HD 15337 b\\
$i$ ($\degree$) & $89.30_{-0.57}^{+0.48}$\\
$e$ & $0.058_{-0.016}^{+0.022}$\\
$\omega$ ($\degree$) & $71.62_{-40.30}^{+52.51}$\\
$R_\mathrm{b}$ ($R_\oplus$) & $1.770_{-0.030}^{+0.032}$\\
$M_\mathrm{b}$ ($M_\oplus$) & $6.519_{-0.400}^{+0.409}$\\
$\rho_\mathrm{b}$ (g\,cm$^{-3}$) & $6.458_{-0.510}^{+0.518}$\\
\hline

Model parameters of HD 15337 c\\
$R_\mathrm{c}/R_\star$ & $0.02707_{-0.00077}^{+0.00091}$\\
$P_\mathrm{c}$ (days) & $17.180546_{-0.000026}^{+0.000021}$\\
$T_\mathrm{0,c}$ (BJD) & $2458414.55162_{-0.0014}^{+0.0015}$\\
$e_\mathrm{c}\cos{\omega_\mathrm{c}}$ & $0.017_{-0.089}^{+0.082}$\\
$e_\mathrm{c}\sin{\omega_\mathrm{c}}$ & $0.048_{-0.055}^{+0.053}$\\
$\cos{i_\mathrm{c}}$ &  $0.0278_{-0.0013}^{+0.0012}$\\
$K_\mathrm{c}$ (km\,s$^{-1}$) & $0.00192_{-0.00032}^{+0.00036}$\\
\hline

Derived parameters of HD 15337 c\\
$i$ ($\degree$) & $88.41\pm0.07$\\
$e$ & $0.096_{-0.045}^{+0.059}$\\
$\omega$ ($\degree$) &  $56.63_{-63.30}^{+77.89}$\\
$R_\mathrm{c}$ ($R_\oplus$) & $2.526_{-0.075}^{+0.086}$\\
$M_\mathrm{c}$ ($M_\oplus$) & $6.792_{-1.143}^{+1.302}$\\
$\rho_\mathrm{c}$ (g\,cm$^{-3}$) & $2.303_{-0.414}^{+0.505}$\\
\hline

Additional Model Parameters\\
$\Delta RV_\mathrm{HARPS\_1}$ (km\,s$^{-1}$) & $0.0187_{-0.0019}^{+0.0020}$\\
$\Delta RV_\mathrm{PFS}$ (km\,s$^{-1}$) & $3.8137\pm0.0021$\\
$P_\mathrm{sin}$ (days) & $19557_{-2612}^{+3425}$\\
$K_\mathrm{sin}$ (km\,s$^{-1}$) & $0.0177_{-0.0038}^{+0.0044}$\\

\hline

\end{tabular}
\end{table}

\begin{figure*}
     \centering
        \includegraphics[width=\linewidth]{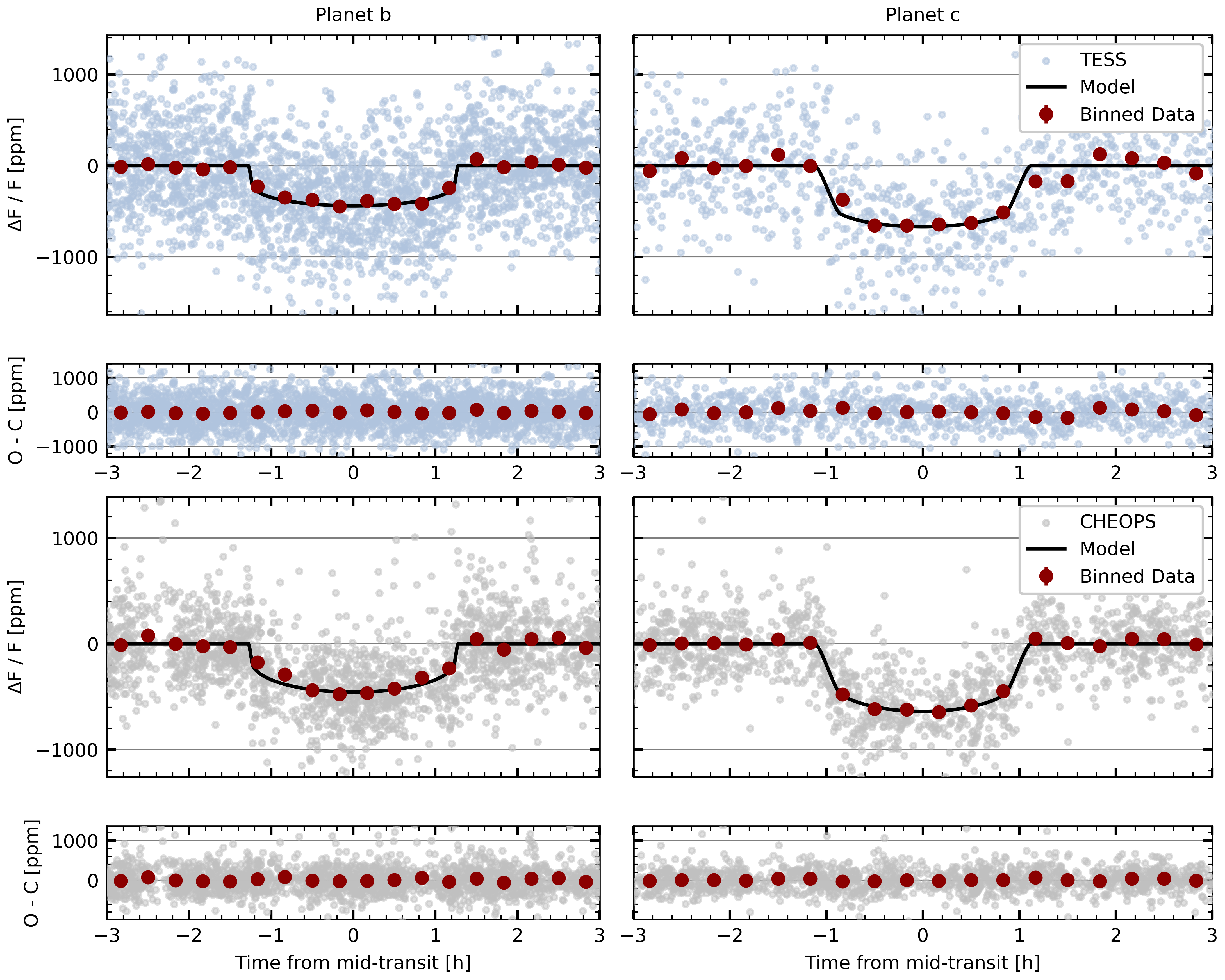}
        \caption{Phase-folded light curves from TESS (upper panels) and CHEOPS (lower panels) measurements. The best-fit model of the transit is plotted in black for HD 15337 b (left) and HD 15337 c (right). The bottom panels for each instrument show the residuals of the best-fit model. We over-plot the binned light curves and residuals with 20-min bins.}
        \label{transitfit}
\end{figure*}

\begin{figure*}[h]
     \centering
        \includegraphics[width=0.90\linewidth]{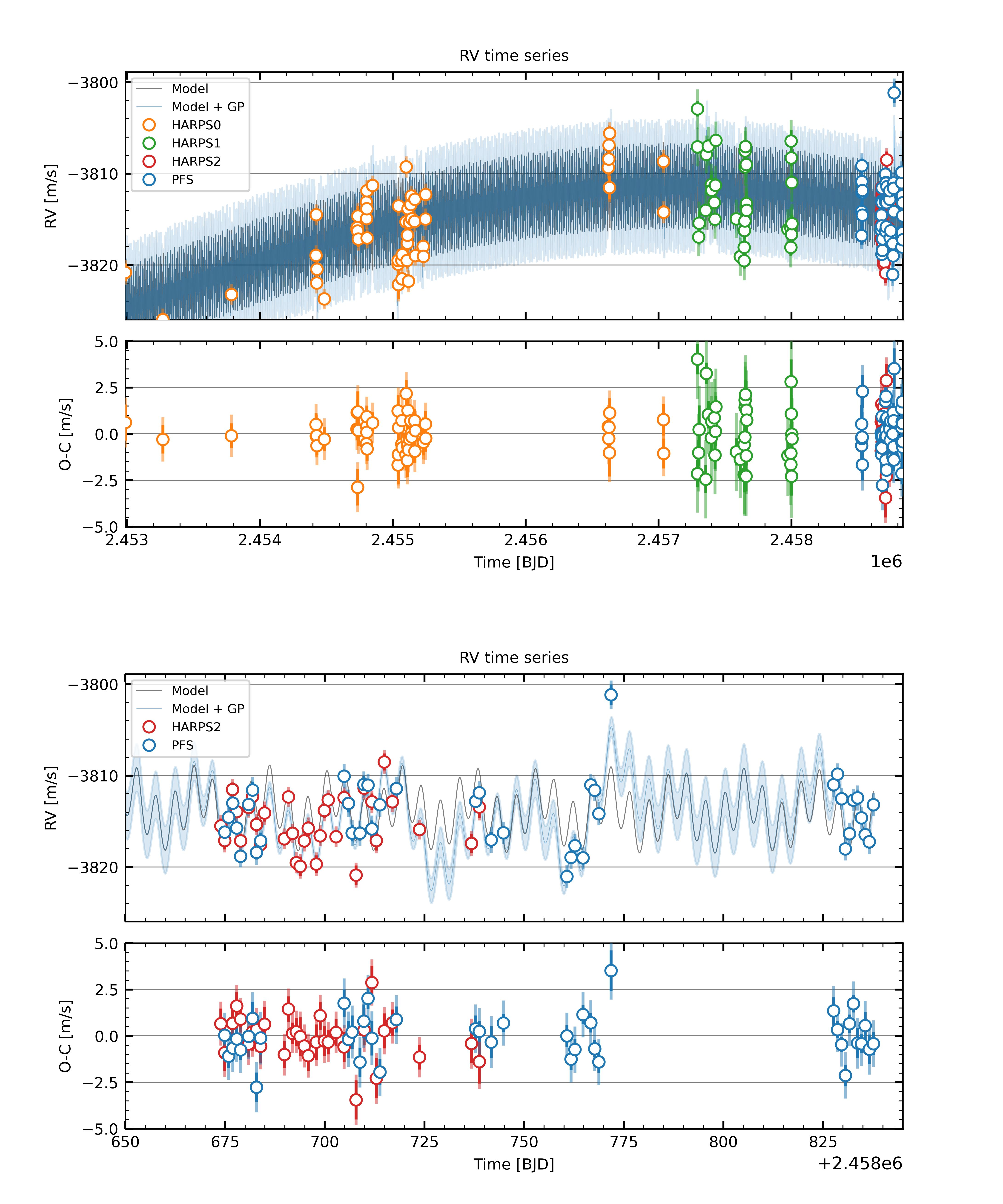}
        \caption{Full RV time series for HD 15337 corrected for instrumental variability. The best-fit model with both Keplerian signals and the model with the stellar activity GP are over-plotted in blue and light blue respectively. \textit{Bottom panels}: zoom of the most recent HARPS and PFS data.}
        \label{rvfull}
\end{figure*}

\begin{figure*}
     \centering
        \includegraphics[width=\linewidth]{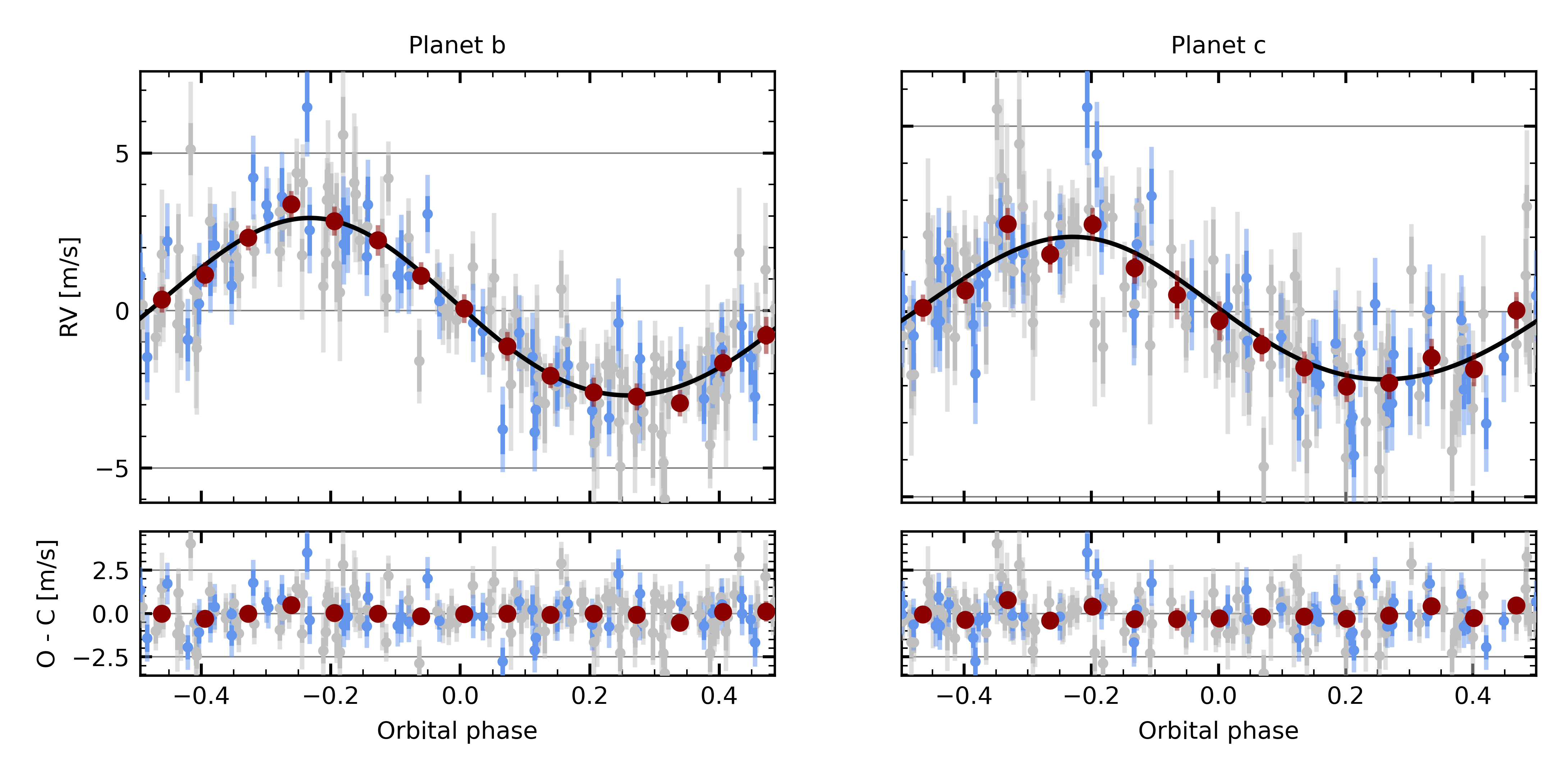}
        \caption{Phase-folded radial velocity curves (top) from HARPS (grey dots) and PFS (blue dots) measurements. The best-fit model of the Keplerian signal is plotted in black for HD 15337 b (left) and HD 15337 c (right). The bottom panel shows the residuals of the best-fit model like in Figure \ref{transitfit}. We over-plot the binned data and residuals.}
        \label{rvfit}
\end{figure*}

\begin{figure*}
     \centering
        \includegraphics[width=\linewidth]{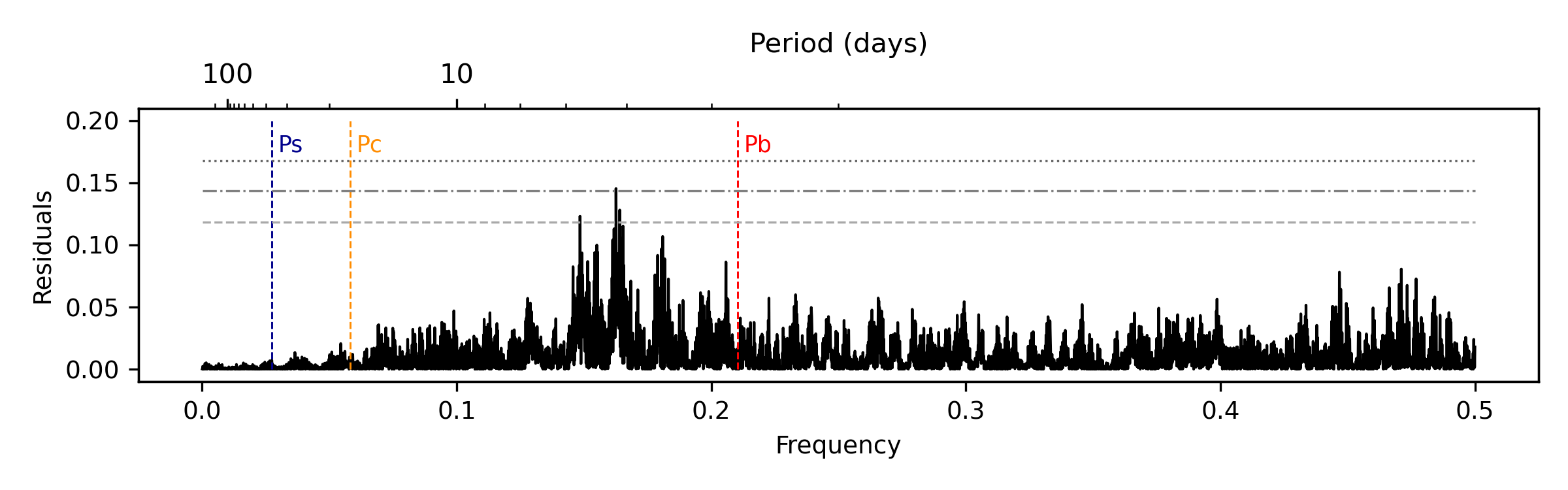}
        \caption{Periodogram of the residuals from the RV fit. There are no longer any peaks at the orbital periods of the planet or the stellar rotation period, showing a good fit from the GP and the Keplerian models. Horizontal lines show the FAP thresholds as in Figure \ref{period_init}.}
        \label{glsresi}
\end{figure*}

The best-fit value of each of the parameters included in the joint analysis is obtained from the median of the posterior distribution, with the associated error corresponding to the 1$\sigma$ confidence interval. A summary of the best-fit results can be found in Table \ref{tableresults} for the most relevant system and planet parameters, and the full results, including the posteriors of the detrending parameters and GP hyperparameters are shown in Table \ref{priors}. Figure \ref{transitfit} shows the phase folded transits from TESS and CHEOPS with the best-fit transit model, while Figures \ref{rvfull} and \ref{rvfit} show the full HARPS and PFS RV time series and the phase folded RVs for both planet models, respectively. Figure \ref{glsresi} shows the periodogram of the residuals of the RV fit, showing no peaks remaining near the planet periods or the stellar rotation period. All significant peaks shown in Figures \ref{period_init} and \ref{period_planets} that are not caused by orbiting planets were successfully removed by our GP model.\par

HD 15337 b has a measured radius of $R_\mathrm{b}=1.770_{-0.030}^{+0.032}\ R_\oplus$ and a mass of $M_\mathrm{b}=6.519_{-0.400}^{+0.409}\ M_\oplus$, corresponding to a mean density of $\rho_\mathrm{b}=6.458_{-0.510}^{+0.518}$ g\,cm$^{-3}$, consistent with a rocky planet with a thin atmosphere. For HD 15337 c, we obtain a radius of $R_\mathrm{c}=2.526_{-0.086}^{+0.075}\ R_\oplus$ and a mass of $M_\mathrm{c}=6.792_{-1.143}^{+1.302}\ M_\oplus$, which gives a mean density of $\rho_\mathrm{c}=2.303_{-0.414}^{+0.505
}$ g\,cm$^{-3}$. 

\subsection{Internal structure model}\label{intstruct}
Using the results from the joint analysis, we model the internal structure of HD 15337 b and c. We follow the method described in \cite{Leleu2021}, which is based on \cite{Dorn2017}. In the following, we briefly summarise the most important aspects of the model. Our internal structure model assumes each planet to be fully spherically symmetric and to consist of four fully distinct layers: an iron core modelled using equations of state from \cite{Hakim2018}, a silicate mantle \citep{Sotin2007}, a water layer \citep{Haldemann2020} and a pure H/He atmosphere as described in \cite{LopezFortney2014}. We further assume that the Si/Mg/Fe ratios of each planet match the ones of the host star exactly \citep{Thiabaud2015}. \par
Our Bayesian inference model uses both stellar and planetary observables as input parameters, more specifically the age, mass, radius, effective temperature and abundances of the star and the transit depth, period and mass relative to the star for each planet. We assume a uniform prior for the mass fractions of the innermost three layers (iron core, mantle and water), with the additional conditions that the sum of the three mass fractions is always one and the upper limit of the water mass fraction is 0.5 \citep{Thiabaud2014, Marboeuf2014}. For the mass of the H/He layer, we choose a prior that is uniform in log. Because of the intrinsic degeneracy of the problem, the results of the model do depend on the chosen priors and would differ if very different priors were chosen. The priors used in our analysis are provided in Table \ref{priors_int} and, in section \ref{discussion}, we discuss how different choices of priors and different model's assumptions would affect our results. Furthermore, we model both planets simultaneously. \par
\begin{figure}
	\centering
	\includegraphics[width=0.5\textwidth]{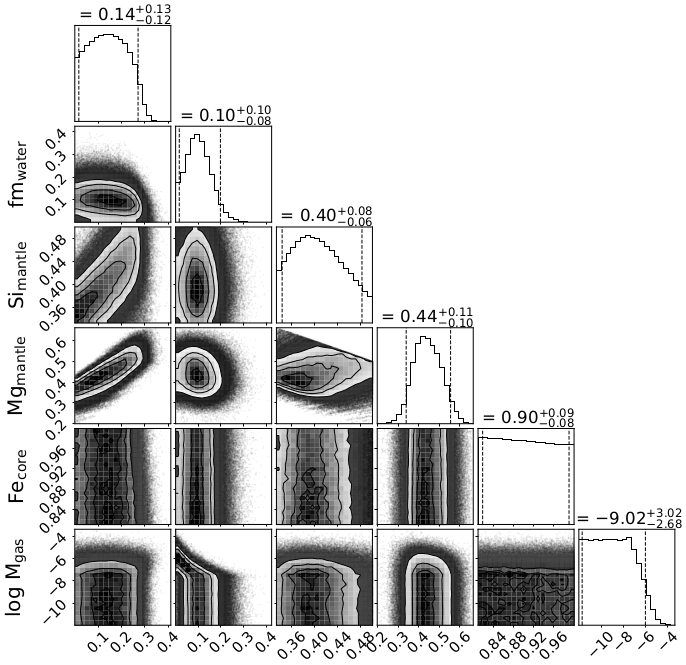}
	\caption{Posterior distribution of the most important internal structure parameters for HD 15337 b. The parameters depicted are the mass fractions of the inner iron core and water layer with respect to the solid part of the planet (which together with the mass fraction of the silicate mantle add up to 1), the molar fractions of Si and Mg in the mantle layer and Fe in the inner iron core and the logarithm with base 10 of the total mass of the H/He layer in Earth masses. The values and errors on top of each column correspond to the median and the 5$^\textrm{th}$ and 95$^\textrm{th}$ percentile.}
	\label{internalstructure_b}
\end{figure}
\begin{figure}
	\centering
	\includegraphics[width=0.5\textwidth]{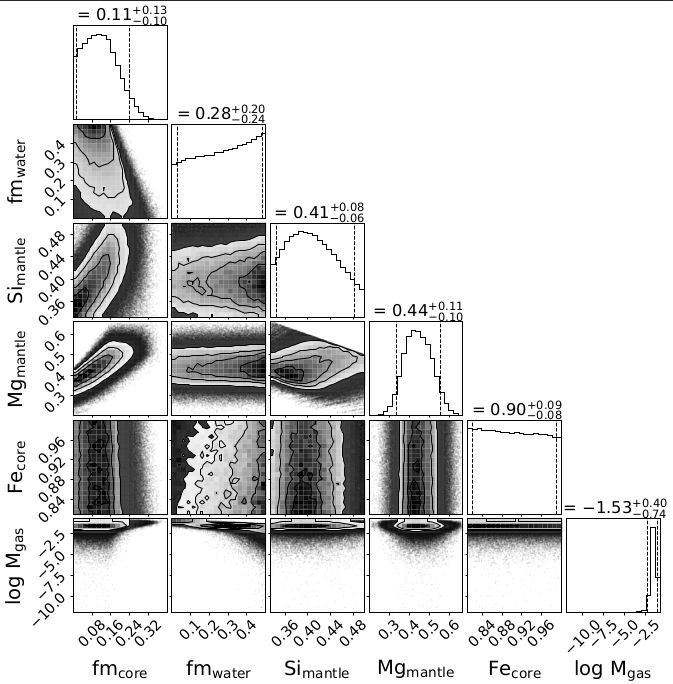}
	\caption{Same as Figure \ref{internalstructure_b} but for HD 15337 c.}
	\label{internalstructure_c}
\end{figure}

The resulting posterior distributions from our internal structure analysis for HD 15337 b and c are shown in Figures \ref{internalstructure_b} and \ref{internalstructure_c}, where the errors correspond to the 5$^\textrm{th}$ and 95$^\textrm{th}$ percentile. We find a core mass fraction (fm$_\mathrm{core}$) of 0.14$^{+0.13}_{-0.12}$ for planet b and 0.11$^{+0.13}_{-0.10}$ for planet c. While our models show that the mass of H/He (m$_\mathrm{gas}$) in HD 15337 b is negligibly small, the planet may host a water layer, as the mass fraction with respect to the solid part of the planet (fm$_\mathrm{water}$) is 0.10$^{+0.10}_{-0.08}$. Conversely, for HD 15337 c the water mass fraction is almost completely unconstrained at 0.28$^{+0.20}_{-0.24}$, while the planet seems to host a significant H/He layer with a mass of 0.03$^{+0.04}_{-0.02}$ $M_{\oplus}$ and a thickness of 0.60$^{+0.18}_{-0.29}$ $R_{\oplus}$. The posteriors of the internal structure best-fit model are summarised in Table \ref{priors_int}.


\section{Discussion}
\label{discussion}

The high precision photometry measurements from CHEOPS allow us to improve the radius precision of HD 15337 b from the $\sim3.5\%$ from \cite{gandolfi2019} and \cite{dumusque2019hot} to 1.8\%, turning it into one of the highest precision radius measurements for super-Earths. HD 15337 c also improved its radius precision to  3\%. With this, both planets are now under the 3\% threshold for radius precision, which is key to improve the constraint on the water mass fraction \citep{Dorn2017}. We also significantly improve the mass precision of HD 15337 b with the help of the new HARPS and PFS data. Measured at around 11-13\% by \cite{gandolfi2019} and \cite{dumusque2019hot}, the mass uncertainty from our analysis is around half those values, sitting at 6.5\%. While there was an improvement in the mass uncertainty of HD 15337 c, we were only able to reach a precision of 17\%, very similar to the 19-21\% obtained by the mentioned authors. We note that the median value of our masses is lower on both planets than what was measured before. \cite{gandolfi2019} reports a mass of $M_\mathrm{b}=7.51_{-1.01}^{+1.09}\ M_\oplus$ for HD 15337 b, with \cite{dumusque2019hot} providing a similar value and uncertainty at $M_\mathrm{b}=7.20\pm0.81\ M_\oplus$. For HD 15337 c, the difference is larger but still within the 1$\sigma$ range of the result of \cite{dumusque2019hot}, $M_\mathrm{c}=8.79\pm1.68\ M_\oplus$. As expected, the new RV measurements from HARPS and PFS allow a stronger constrain of the long-term trends as well as the stellar activity signals by expanding the temporal range, which in turn allows for a better fit of the RV signal of both planets, leading to a more accurate result. We can also see that while the masses of both planets are similar, HD 15337 c has a radius that is $\sim1.4$ times larger than HD 15337 b. This results in a lower density, suggesting a planet with a more significant water layer and gas envelope.\par

The long-term trend shown in the RV time series is most likely caused by stellar activity as mentioned before, due to the fact that the same type of trend is also present in the FWHM indicator. Alternatively, it is possible that the signal is being caused by a long-period companion. A stellar companion could explain the trend seen in FWHM if the radial velocity of the companion is close enough to the target star. However, for a planetary companion, the FWHM trend would have to be due to the instrument and the correlation with the RV trend would have to be coincidental.

As stated in section 4.3.2, the trend was modelled with a sinusoidal function, whose best-fit results show a period of $53.6_{-7.2}^{+9.4}$ years and a semi-amplitude of $0.0177_{-0.0038}^{+0.0044}$ km\,s$^{-1}$, an order of magnitude higher than the RV signature of the two planets. If we assume this trend is caused by a long-period companion, we estimate its minimum mass to be $M\sin{i}=4\ M_\mathrm{Jup}$. It would be theoretically possible that this trend was caused by the stellar companion described in section 3, but we find that with a period of 540 yr \citep{Lester21}, the orbit would need to be eccentric to produce such a trend in the time span of the HARPS observations ($\sim$16 yrs). We estimated the amplitude of the signal caused by the stellar companion, given the mass ratio of 0.14 from \cite{Lester21}. This is one order of magnitude higher than the amplitude we found in the current RVs, which could be the case for an eccentric orbit. Therefore, it is unclear that this companion is the cause of the RV trend. Initially, this trend was modelled as a second-degree polynomial as in \cite{gandolfi2019}, which provides consistent results (at 1$\sigma$) with the sinusoidal model. We decided to adopt the sinusoidal model as it provides a more direct constraint on the hypothesis of a gravitationally bound companion.\par

The periodogram of the RV residuals (Figure \ref{glsresi}) shows some hints at the periods of 6.73, 6.15 and 5.53 days. These peaks were not detected before the removal of the planetary and stellar activity signals and are also not present in the periodograms of the indicators. However, since they are below the 0.1\% FAP level, there is not enough evidence of a statistically significant signal at these periods \citep{baluev2008assessing}. They also do not correspond to aliases of the stellar rotation period and is unlikely that they are of a planetary nature, given the proximity to the orbit of HD 15337 b and the fact that there are no hints of a transit around those periods in the TESS light curves. As such, the nature of these peaks is currently unknown.

\begin{figure}
	\centering
	\includegraphics[width=0.5\textwidth]{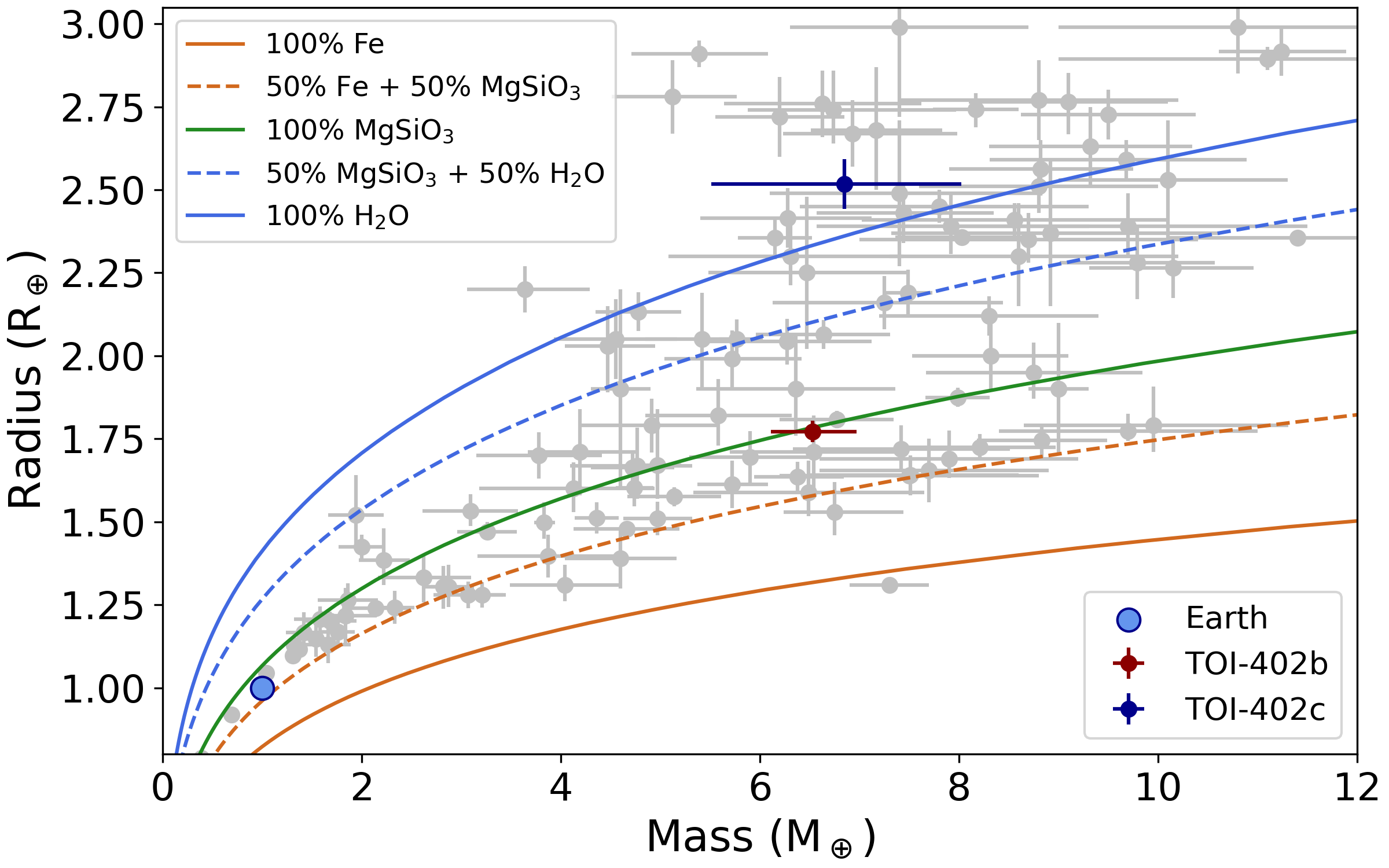}
	\caption{
 Mass-radius plot showing the position of HD 15337 b and HD 15337 c compared to other small exoplanets. 
 }
	\label{mrdiagram}
\end{figure}

Figure \ref{mrdiagram} shows the position of HD 15337 b and HD 15337 c in the mass-radius diagram compared to other small planets with a mass precision <25\%. The composition models from \cite{zeng2016mass} are overplotted ranging from a full iron composition to a full water world, to provide a general view of the position of each planet in the field. HD 15337 b is located on the upper range of the super-Earth population below the radius gap \citep{fulton2017california}, lying close to the pure silicate line from the model of \cite{zeng2016mass}. HD 15337 c on the other hand is placed within the sub-Neptune population, on the other side of the radius gap and above the predicted 100\% water line, hinting at the presence of a gas envelope. The results of our internal structure analysis agree with the position of both planets in the MR diagram, as we find a H/He layer in planet c with a significant water layer and we confirm planet b to be in the super-Earth population. Despite laying near the pure silicate line according to the models from \cite{zeng2016mass}, HD 15337 b shows a relatively high iron mass fraction in both core and mantle with our model.\par

Our internal structure model includes liquid phases in the EOS of the inner core and the water layer, but we fix the temperature and pressure to 300 K and 1 bar outside the water layer and model the gas layer separately from the rest of the planet. As HD 15337 b orbits close to its star, the high irradiation causes it to reach high equilibrium temperatures ($\sim$1000 K). At these temperatures, it is highly unlikely that the water is fully condensed, and with a density that is below Earth's density, this planet may have a molten mantle with water present in it. To tackle this problem, \cite{dorn2021hidden} presents an internal structure model that takes into account this aspect and shows that the presence of water in the mantle can lead to a significant underestimation of the water mass fraction given the same observed mass and radius. Determining the water content of exoplanets with accuracy is particularly relevant to the search for habitable worlds and can affect dynamic and structural properties like the differentiation between core and mantle \citep{bonati2021structure} and atmospheric composition \citep{gaillard2021diverse}. For a planet with the mass and radius of HD 15337 b, \cite{dorn2021hidden} predicts a water mass fraction 15-20\% higher than obtained with our model. It should be noted that this difference appears to be smaller for more massive planets. \par
Another assumption we make in our model is that the abundances of the planets are exactly the same as the host star. However, \cite{adibekyan2021compositional} shows that while there is a correlation, it is unlikely to be a 1:1 relationship, and the ratios of Si/Mg/Fe may not necessarily match the ones from the star. We performed a new run of the internal structure model for both planets without constraining the abundances to compare the results. We saw that without the priors in the abundances, both planets show a similar range of values in the posterior of the core mass fraction. However, the distribution indicates that smaller mass fractions are accepted more often than before. The same is true for the water mass fraction of HD 15337 b, with the posterior being indistinguishable from zero at a 2$\sigma$ confidence level. It is, therefore, more likely that there is not a significant water layer on this planet, compared to what the model with constrained abundances showed. We also note an increase in the range in the mantle mass fraction, compensating for the smaller core and water fractions. The mass of the gas envelope of HD 15337 c does not show significant changes. \par
It has been suggested that HD 15337 b could host a secondary atmosphere, after losing its initial H/He envelope to photo evaporation \citep{dumusque2019hot}. Confirming this would require further research into the evolution of these planets and the real composition of the gas envelope present in HD 15337 c. \par
We provide improved limits on radius, mass and internal structure and composition parameters that should provide a strong basis for follow-up studies. HD 15337 is expected to be observed once again, this time by JWST, which could further improve these limits and potentially provide enough data to better constrain the evolution and formation of this system.

\section{Summary}

In this paper, we use all available measurements of HD 15337 obtained from CHEOPS, HARPS and PFS to perform a joint fit of the transits and RVs of the HD 15337 system. We significantly improve the precision of the radius and mass of HD 15337 b, to 1.7\% and 6.2\% respectively. We also improve the precision of the radius of HD 15337 c to 3\%, but we were unable to decrease the mass uncertainties despite the additional RV points, sitting at $\sim$18\%. With our internal structure model, fitting both planets at the same time, we find that HD 15337 b is most likely a rocky planet and that while our calculations show a water mass fraction higher than zero at a 2$\sigma$ level, the planet is also compatible with a dry composition. We find that while the water mass fraction of HD 15337 c remains unconstrained, the model shows a gaseous envelope with a mass higher than 0.01 $M_{\oplus}$. We conclude that the results of the internal structure analysis are highly dependent on the model and the assumptions used.

\footnotetext{https://exoplanetarchive.ipac.caltech.edu, retrieved on April 27, 2023.}


\begin{acknowledgements}
CHEOPS is an ESA mission in partnership with Switzerland with important contributions to the payload and the ground segment from Austria, Belgium, France, Germany, Hungary, Italy, Portugal, Spain, Sweden, and the United Kingdom. The CHEOPS Consortium would like to gratefully acknowledge the support received by all the agencies, offices, universities, and industries involved. Their flexibility and willingness to explore new approaches were essential to the success of this mission. 
This work was supported by FCT - Fundação para a Ciência e a Tecnologia through national funds by these grants: UIDB/04434/2020, UIDP/04434/2020, 2022.06962.PTDC, 2022.04416.PTDC.
N.M.R. acknowledges support from FCT through grant DFA/BD/5472/2020.
O.D.S.D. is supported in the form of work contract (DL 57/2016/CP1364/CT0004) funded by national funds through FCT. 
S.C.C.B. acknowledges support from FCT through FCT contracts nr. IF/01312/2014/CP1215/CT0004. 
DG gratefully acknowledges financial support from the CRT foundation under Grant No. 2018.2323 ``Gaseousor rocky? Unveiling the nature of small worlds''. 
JAEg and YAl acknowledge support from the Swiss National Science Foundation (SNSF) under grant 200020\_192038. 
LMS gratefully acknowledges financial support from the CRT foundation under Grant No. 2018.2323 ‘Gaseous or rocky? Unveiling the nature of small worlds’. 
This work has been carried out within the framework of the NCCR PlanetS supported by the Swiss National Science Foundation under grants 51NF40\_182901 and 51NF40\_205606. 
TWi and ACCa acknowledge support from STFC consolidated grant numbers ST/R000824/1 and ST/V000861/1, and UKSA grant number ST/R003203/1. 
The Belgian participation to CHEOPS has been supported by the Belgian Federal Science Policy Office (BELSPO) in the framework of the PRODEX Program, and by the University of Liège through an ARC grant for Concerted Research Actions financed by the Wallonia-Brussels Federation. 
L.D. is an F.R.S.-FNRS Postdoctoral Researcher. 
NCSa acknowledges funding by the European Union (ERC, FIERCE, 101052347). Views and opinions expressed are however those of the author(s) only and do not necessarily reflect those of the European Union or the European Research Council. Neither the European Union nor the granting authority can be held responsible for them. 
S.G.S. acknowledge support from FCT through FCT contract nr. CEECIND/00826/2018 and POPH/FSE (EC). 
V.A. is supported by FCT through national funds by the following grant: 2022.06962.PTDC (http://doi.org/10.54499/2022.06962.PTDC).
RAl, DBa, EPa, and IRi acknowledge financial support from the Agencia Estatal de Investigación of the Ministerio de Ciencia e Innovación MCIN/AEI/10.13039/501100011033 and the ERDF “A way of making Europe” through projects PID2019-107061GB-C61, PID2019-107061GB-C66, PID2021-125627OB-C31, and PID2021-125627OB-C32, from the Centre of Excellence “Severo Ochoa'' award to the Instituto de Astrofísica de Canarias (CEX2019-000920-S), from the Centre of Excellence “María de Maeztu” award to the Institut de Ciències de l’Espai (CEX2020-001058-M), and from the Generalitat de Catalunya/CERCA programme. 
XB, SC, DG, MF and JL acknowledge their role as ESA-appointed CHEOPS science team members. 
LBo, VNa, IPa, GPi, RRa, and GSc acknowledge support from CHEOPS ASI-INAF agreement n. 2019-29-HH.0.
P.E.C. is funded by the Austrian Science Fund (FWF) Erwin Schroedinger Fellowship, program J4595-N.
ABr was supported by the SNSA. 
This project was supported by the CNES. 
B.-O. D. acknowledges support from the Swiss State Secretariat for Education, Research and Innovation (SERI) under contract number MB22.00046. 
This project has received funding from the European Research Council (ERC) under the European Union’s Horizon 2020 research and innovation programme (project {\sc Four Aces}, grant agreement No 724427). It has also been carried out in the frame of the National Centre for Competence in Research PlanetS supported by the Swiss National Science Foundation (SNSF). DE acknowledges financial support from the Swiss National Science Foundation for project 200021\_200726.
MF and CMP gratefully acknowledge the support of the Swedish National Space Agency (DNR 65/19, 174/18). 
M.G. is an F.R.S.-FNRS Senior Research Associate. 
MNG is the ESA CHEOPS Project Scientist and Mission Representative, and as such also responsible for the Guest Observers (GO) Programme. MNG does not relay proprietary information between the GO and Guaranteed Time Observation (GTO) Programmes, and does not decide on the definition and target selection of the GTO Programme. 
CHe acknowledges support from the European Union H2020-MSCA-ITN-2019 under Grant Agreement no. 860470 (CHAMELEON).
SH gratefully acknowledges CNES funding through the grant 837319. 
KGI is the ESA CHEOPS Project Scientist and is responsible for the ESA CHEOPS Guest Observers Programme. She does not participate in, or contribute to, the definition of the Guaranteed Time Programme of the CHEOPS mission through which observations described in this paper have been taken, nor to any aspect of target selection for the programme.
K.W.F.L. was supported by Deutsche Forschungsgemeinschaft grants RA714/14-1 within the DFG Schwerpunkt SPP 1992, Exploring the Diversity of Extrasolar Planets. 
This work was granted access to the HPC resources of MesoPSL financed by the Region Ile de France and the project Equip@Meso (reference ANR-10-EQPX-29-01) of the programme Investissements d'Avenir supervised by the Agence Nationale pour la Recherche.
ML acknowledges support of the Swiss National Science Foundation under grant number PCEFP2\_194576.
PM acknowledges support from STFC research grant number ST/M001040/1. 
This work was also partially supported by a grant from the Simons Foundation (PI Queloz, grant number 327127).
GyMSz acknowledges the support of the Hungarian National Research, Development and Innovation Office (NKFIH) grant K-125015, a a PRODEX Experiment Agreement No. 4000137122, the Lend\"ulet LP2018-7/2021 grant of the Hungarian Academy of Science and the support of the city of Szombathely.
V.V.G. is an F.R.S-FNRS Research Associate. 
NAW acknowledges UKSA grant ST/R004838/1.
AN and JV acknowledge support from the Swiss National Science Foundation (SNSF) under grant PZ00P2\_208945.
\end{acknowledgements}


\bibliographystyle{aa} 
\bibliography{refs} 


\onecolumn

\begin{appendix}
\section{Additional Tables}

\begin{table*}[h]
\small
\centering
\caption{Priors used for the joint transit and RV modelling and corresponding posteriors from the best-fit model.}
\label{priors}
\begin{tabular}{lcc}\hline
Parameters & Prior & Best-fit\\ \hline
Transit model parameters\\
$\rho_\star$ ($\rho_\sun$) & $\mathcal{N}(1.344, 0.076)$ & $1.334_{-0.079}^{+0.075}$\\
$u_{1,\mathrm{TESS}}$ & $\mathcal{N}(0.47621, 0.00145)$ & $0.4761_{-0.0012}^{+0.0015}$\\
$u_{2,\mathrm{TESS}}$ & $\mathcal{N}(0.12172, 0.00220)$ & $0.1219_{-0.0027}^{+0.0024}$\\
$u_{1,\mathrm{CHEOPS}}$ & $\mathcal{N}(0.61973, 0.00133)$ & $0.6198_{-0.0012}^{+0.0013}$\\
$u_{2,\mathrm{CHEOPS}}$ & $\mathcal{N}(0.0696, 0.00182)$ & $0.0696_{-0.0019}^{+0.0018}$\\
\hline

RV model parameters\\
$v_\mathrm{0}$ (km\,s$^{-1}$) & $\mathcal{U}(-4.0, 0.0)$ & $-3.8292_{-0.0042}^{+0.0033}$\\
\hline

Model parameters of HD 15337 b\\
$R_\mathrm{b}/R_\star$ & $\mathcal{U}(0.01, 0.03)$ & $0.01898_{-0.00026}^{+0.00030}$\\
$P_\mathrm{b}$ (days) & $\mathcal{U}(4.7555, 4.7565)$ & $4.7559804_{-0.0000037}^{+0.0000062}$\\
$T_\mathrm{0,b}$ (BJD-2457000) & $\mathcal{U}(1411.457, 1411.467)$\tablefootmark{a} & $1411.4623_{-0.0013}^{+0.0008}$\\
$e_\mathrm{b}$ & $\beta(0.867, 3.03)$ & $0.058_{-0.016}^{+0.022}$\\
$b_\mathrm{b}$ & $\mathcal{U}(0.0, 2.0)$ & $0.17_{-0.11}^{+0.13}$\\
$K_\mathrm{b}$ (km\,s$^{-1}$) & $\mathcal{U}(0.0, 0.01)$ & $0.00282\pm0.00015$\\
\hline

Model parameters of HD 15337 c\\
$R_\mathrm{c}/R_\star$ & $\mathcal{U}(0.01, 0.04)$ & $0.02707_{-0.00077}^{+0.00091}$\\
$P_\mathrm{c}$ (days) & $\mathcal{U}(17.180, 17.181)$ & $17.180546_{-0.000026}^{+0.000021}$\\
$T_\mathrm{0,b}$ (BJD-2457000) & $\mathcal{U}(1414.545, 1414.555)$\tablefootmark{a} & $2458414.55162_{-0.0014}^{+0.0015}$\\
$e_\mathrm{c}$ & $\beta(0.867, 3.03)$ & $0.096_{-0.045}^{+0.059}$\\
$b_\mathrm{c}$ & $\mathcal{U}(0.0, 2.0)$ & $0.89_{-0.07}^{+0.08}$\\
$K_\mathrm{c}$ (km\,s$^{-1}$) & $\mathcal{U}(0.0, 0.01)$ & $0.00192_{-0.00032}^{+0.00036}$\\
\hline

Additional RV model parameters\\
$\Delta RV_\mathrm{HARPS\_1}$ (km\,s$^{-1}$) & $\mathcal{N}(0.022, 0.01)$ & $0.0187_{-0.0019}^{+0.0020}$\\
Jitter $\sigma_\mathrm{HARPS\_0}$ & $\mathcal{U}(0.0, 0.0035)$ & $0.0009_\pm0.0003$\\
Jitter $\sigma_\mathrm{HARPS\_1}$ & $\mathcal{U}(0.0, 0.0035)$ & $0.0020_{-0.0004}^{+0.0005}$\\
Jitter $\sigma_\mathrm{HARPS\_2}$ & $\mathcal{U}(0.0, 0.0035)$ & $0.0009_{-0.0003}^{+0.0004}$\\
$\Delta RV_\mathrm{PFS}$ (km\,s$^{-1}$) & $\mathcal{N}(3.815, 0.01)$ & $3.8137\pm0.0021$\\
Jitter $\sigma_\mathrm{PFS}$ & $\mathcal{U}(0.0, 0.0035)$ & $0.0011_\pm0.0002$\\
$P_\mathrm{sin}$ (days) & $\mathcal{U}(5500, 170000)$ & $19557_{-2612}^{+3425}$\\
$K_\mathrm{sin}$ (km\,s$^{-1}$) & $\mathcal{U}(0, 0.025)$ & $0.0177_{-0.0038}^{+0.0044}$\\
$\log{R'_\mathrm{HK,0}}$ & $\mathcal{N}(-4.95, 0.2)$ & $-4.999\pm0.023$\\
$\log{R'_\mathrm{HK,drift1}}$ (d$^{-1}$) & $\mathcal{N}(0.0, 0.2)$ & $8.790_{-2.074}^{+1.810}\times 10^{-5}$\\
$\log{R'_\mathrm{HK,drift2}}$ (d$^{-2}$) & $\mathcal{N}(0.0, 0.2)$ & $-1.910_{-0.366}^{+0.405}\times 10^{-8}$\\
\hline

Additional transit model parameters\\
$\Delta F_\mathrm{CHEOPS}$ (visit 1) & $\mathcal{N}(0.0, 0.001)$ &  $2.188_{-0.185}^{+0.170}\times 10^{-4}$\\
$\Delta F_\mathrm{CHEOPS}$ (visit 2) & $\mathcal{N}(0.0, 0.001)$ & $1.263_{-0.102}^{+0.136}\times 10^{-4}$\\
$\Delta F_\mathrm{CHEOPS}$ (visit 3) & $\mathcal{N}(0.0, 0.001)$ & $1.375_{-0.114}^{+0.131}\times 10^{-4}$\\
$\Delta F_\mathrm{CHEOPS}$ (visit 4) & $\mathcal{N}(0.0, 0.001)$ & $7.343_{-1.138}^{+1.306}\times 10^{-5}$\\
$\Delta F_\mathrm{CHEOPS}$ (visit 5) & $\mathcal{N}(0.0, 0.001)$ & $1.704\pm0.129\times 10^{-4}$\\
$\Delta F_\mathrm{CHEOPS}$ (visit 6) & $\mathcal{N}(0.0, 0.001)$ & $1.129_{-0.130}^{+0.135}\times 10^{-4}$\\
Jitter $\sigma_\mathrm{CHEOPS}$ (all visits) & $\mathcal{U}(0.0, 0.0015)$ & $2.085_{-0.056}^{+0.052}\times 10^{-4}$\\
$\Delta F_\mathrm{TESS}$ (sector 3) & $\mathcal{N}(0.0, 0.001)$ & $3.035_{-1.637}^{+1.478}\times 10^{-5}$\\
$\Delta F_\mathrm{TESS}$ (sector 4) & $\mathcal{N}(0.0, 0.001)$ & $-0.514_{-1.081}^{+1.256}\times 10^{-5}$\\
$\Delta F_\mathrm{TESS}$ (sector 30) & $\mathcal{N}(0.0, 0.001)$ & $-3.106_{-1.225}^{+1.191}\times 10^{-5}$\\
Jitter $\sigma_\mathrm{TESS}$ (all sectors) & $\mathcal{U}(0.0, 0.002)$ & $1.622_{-0.153}^{+0.161}\times 10^{-4}$\\
\hline

GP hyperparameters\\
$A_\mathrm{RV}$ & $\mathcal{U}(0.0, 0.01)$ &  $0.0026_{-0.0003}^{+0.0004}$\\
$\tau_\mathrm{decay}$ & $\mathcal{U}(10, 200)$ & $40.44_{-10.64}^{+19.19}$\\
$\gamma$ & $\mathcal{U}(0.05, 5.0)$ & $2.66_{-1.63}^{+1.45}$\\
$\ln{P_\mathrm{rot}}$ & $\mathcal{U}(\ln{10}, \ln{200})$ & $3.829_{-0.980}^{+1.018}$\\
$A_\mathrm{\log{R'_\mathrm{HK}}}$ & $\mathcal{U}(0.0, 0.1)$ &  $0.0317_{-0.0036}^{+0.0043}$\\
\hline

\end{tabular}

\tablefoot{Uniform priors are defined as $\mathcal{U}$(min, max), normal priors are defined as $\mathcal{N}(\mu, \sigma)$, where $\mu$ is the median value and $\sigma$ is the standard deviation, and beta priors are defined as $\beta$(a, b), where a and b are the $\beta$ distribution coefficients.\\\\
\tablefoottext{a}{Prior is put on the orbital phase $\phi$ as detailed in section 4.3.1}
}\\

\end{table*}

\begin{table*}[h]
\small
\centering
\caption{Priors used for the internal structure modelling and corresponding posteriors from the best-fit model with error bars corresponding to the 5\% and 95\% percentiles.}
\label{priors_int}
\begin{tabular}{lcc}\hline
Parameters & Prior & Best-fit\\ \hline
HD 15337 b\\
fm$_\mathrm{core,b}$ & $\mathcal{U}(0.0, 1.0)$\tablefootmark{a} & $0.14_{-0.12}^{+0.13}$\\
fm$_\mathrm{mantle,b}$ & $\mathcal{U}(0.0, 1.0)$\tablefootmark{a} & $0.75_{-0.14}^{+0.15}$\\
fm$_\mathrm{water,b}$ & $\mathcal{U}(0.0, 0.5)$\tablefootmark{a} & $0.10_{-0.08}^{+0.10}$\\
$\log{\mathrm{m_{gas,b}}}$ (M$_\oplus$) & $\mathcal{U}(-12,\log{(0.5\ \mathrm{M_b}}))$ & $-9.02_{-2.68}^{+3.02}$\\
Si$_\mathrm{mantle,b}$\tablefootmark{b} & & $0.40_{-0.06}^{+0.08}$\\
Mg$_\mathrm{mantle,b}$\tablefootmark{b} & & $0.44_{-0.10}^{+0.11}$\\
Fe$_\mathrm{core,b}$ & $\mathcal{U}(0.81, 1.00)$ & $0.90_{-0.08}^{+0.09}$\\
\hline

HD 15337 c\\
fm$_\mathrm{core,c}$ & $\mathcal{U}(0.0, 1.0)$\tablefootmark{a} & $0.11_{-0.10}^{+0.13}$\\
fm$_\mathrm{mantle,c}$ & $\mathcal{U}(0.0, 1.0)$\tablefootmark{a} & $0.60_{-0.26}^{+0.19}$\\
fm$_\mathrm{water,c}$ & $\mathcal{U}(0.0, 0.5)$\tablefootmark{a} & $0.28_{-0.24}^{+0.20}$\\
$\log{\mathrm{m_{gas,c}}}$ (M$_\oplus$) & $\mathcal{U}(-12,\log{(0.5\ \mathrm{M_c}}))$ & $-1.53_{-0.74}^{+0.40}$\\
Si$_\mathrm{mantle,c}$\tablefootmark{b} & & $0.41_{-0.06}^{+0.08}$\\
Mg$_\mathrm{mantle,c}$\tablefootmark{b} & & $0.44_{-0.10}^{+0.11}$\\
Fe$_\mathrm{core,c}$ & $\mathcal{U}(0.81, 1.00)$ & $0.90_{-0.08}^{+0.09}$\\
\hline

\end{tabular}

\tablefoot{Uniform priors are defined as $\mathcal{U}$(min, max).\\\\
\tablefoottext{a}{As mentioned in section \ref{intstruct}, the priors on the mass fractions are not independent. Instead, samples are drawn uniformly from the tringular simplex where the sum of the three mass fractions is equal to one.}\\
\tablefoottext{b}{Priors are applied on the Si/Fe and Mg/Fe bulk ratios, which are sampled using a Gaussian prior from the stellar values (see Table \ref{stepartable}).}
}\\

\end{table*}

\end{appendix}

\end{document}